\documentclass[10pt,twocolumn,letterpaper]{article}

\usepackage{cvpr}              

\usepackage{graphicx}
\usepackage{amsmath}
\usepackage{amssymb}
\usepackage{booktabs}
\usepackage{multirow}
\usepackage{listings}
\usepackage{xcolor}
\usepackage[pagebackref,breaklinks,colorlinks]{hyperref}

\usepackage[capitalize]{cleveref}
\crefname{section}{Sec.}{Secs.}
\Crefname{section}{Section}{Sections}
\Crefname{table}{Table}{Tables}
\crefname{table}{Tab.}{Tabs.}

\def\cvprPaperID{7083} 
\def\confName{CVPR}
\def\confYear{2023}

\begin{document}

\title{Learned Image Compression with Mixed Transformer-CNN Architectures}
\author{Jinming Liu$^{14}$ \qquad Heming Sun$^{23}$\thanks{Heming Sun is the corresponding author.} \qquad Jiro Katto$^{12}$ \\
\normalsize{$^1$Department of Computer Science and Communication Engineering, Waseda University, Tokyo, Japan} \\ \normalsize{$^2$Waseda Research Institute for Science and Engineering, Waseda University, Tokyo, Japan} \\ \normalsize{$^3$JST PRESTO, 4-1-8 Honcho, Kawaguchi, Saitama, Japan \qquad $^4$ Shanghai Jiao Tong University, Shanghai, China}
\\ \small{\{jmliu@toki., hemingsun@aoni., katto@\}waseda.jp}
}
\maketitle
\begin{abstract}
Learned image compression (LIC) methods have exhibited promising progress and superior rate-distortion performance compared with classical image compression standards. Most existing LIC methods are Convolutional Neural Networks-based (CNN-based) or Transformer-based, which have different advantages. Exploiting both advantages is a point worth exploring, which has two challenges: 1) how to effectively fuse the two methods? 2) how to achieve higher performance with a suitable complexity? In this paper, we propose an efficient parallel Transformer-CNN Mixture (TCM) block with a controllable complexity to incorporate the local modeling ability of CNN and the non-local modeling ability of transformers to improve the overall architecture of image compression models. Besides, inspired by the recent progress of entropy estimation models and attention modules, we propose a channel-wise entropy model with parameter-efficient swin-transformer-based attention (SWAtten) modules by using channel squeezing. Experimental results demonstrate our proposed method achieves state-of-the-art rate-distortion performances on three different resolution datasets (i.e., Kodak, Tecnick, CLIC Professional Validation) compared to existing LIC methods. The code is at \url{https://github.com/jmliu206/LIC_TCM}.
\end{abstract}

\vspace{-2mm}
\section{Introduction}
\label{sec:intro}
Image compression is a crucial topic in the field of image processing. With the rapidly increasing image data, lossy image compression plays an important role in storing and transmitting efficiently. In the passing decades, there were many classical standards, including JPEG \cite{wallace1992jpeg}, WebP \cite{webp}, and VVC \cite{vvc}, which contain three steps: transform, quantization, and entropy coding, have achieved impressive Rate-Distortion (RD) performance. 
On the other hand, different from the classical standards, end-to-end learned image compression (LIC) is optimized as a whole. Some very recent LIC works \cite{zou2022devil,zhu2022transformerbased,xie2021enhanced, chen2022two,he2022elic,wang2022neural} have outperformed VVC which is the best classical image and video coding standards at present, on both Peak signal-to-noise ratio (PSNR) and Multi-Scale Structural Similarity (MS-SSIM).
This suggests that LIC has great potential for next-generation image compression techniques.
\begin{figure}[t!]
        \centering
        \includegraphics[scale=0.33]{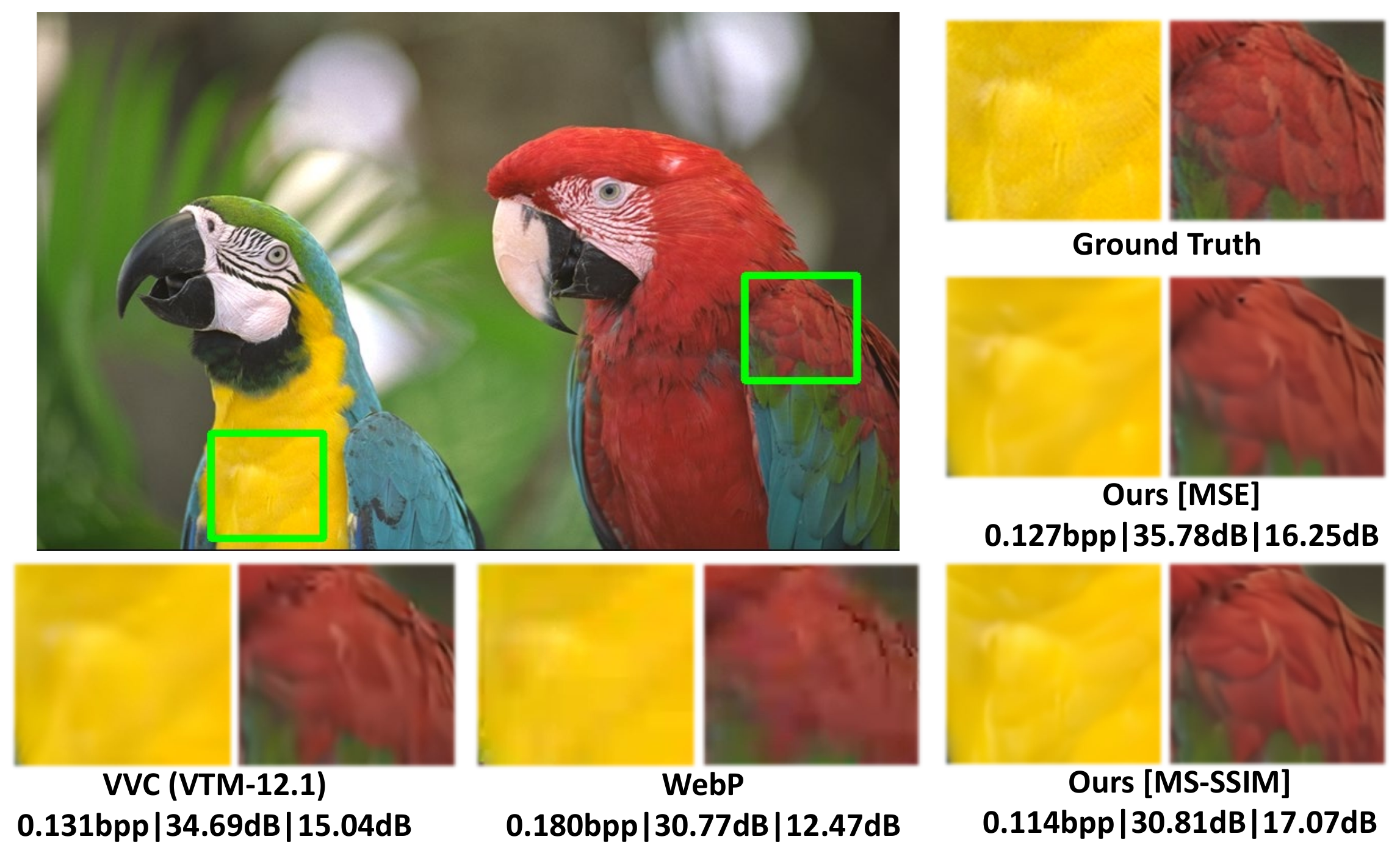}
        \caption{Visualization of decompressed images of $kodim23$ from Kodak dataset based on different methods (The subfigure is titled as ``Method$\mid$Bit rate$\mid$PSNR$\mid$MS-SSIM").}
        \label{fig:v}
        \vspace{-4mm}
\end{figure}
\par
Most LIC methods are CNN-based methods \cite{wu2021learned,fu2021learned,cheng2020learned,liu2019non,lin2023multistage} using the variational auto-encoder (VAE) which is proposed by Ball\'{e} \etal \cite{balle2018variational}. With the development of vision transformers \cite{dosovitskiy2020image,liu2021swin} recently, some vision transformer-based LIC methods \cite{lu2021transformer,zou2022devil,zhu2022transformerbased} are also investigated. For CNN-based example, Cheng \etal \cite{cheng2020learned} proposed a residual block-based image compression model. For transformer-based example, Zou \etal \cite{zou2022devil} tried a swin-transformer-based image compression model. These two kinds of methods have different advantages. CNN has the ability of local modeling, while transformers have the ability to model non-local information. It is still worth exploring whether the advantages of these two methods can be effectively combined with a suitable complexity. In our method, we try to efficiently incorporate both advantages of CNN and transformers by proposing an efficient parallel Transformer-CNN Mixture (TCM) block under a controllable complexity to improve the RD performance of LIC. 
\par
In addition to the type of the neural network, the design of entropy model is also an important technique in LIC. The most common way is to introduce extra latent variables as hyper-prior to convert the probability model of compact coding-symbols to a joint model \cite{balle2018variational}. On that basis, many methods spring up. Minnen \etal \cite{minnen2018joint} utilized the masked convolutional layer to capture the context information. Furthermore, they \cite{minnen2020channel} proposed a parallel channel-wise auto-regressive entropy model by splitting the latent to 10 slices. The results of encoded slices can assist in the encoding of remaining slices in a pipeline manner. 
\par
Recently, many different attention modules \cite{liu2019non,cheng2020learned,zou2022devil} were designed and proposed to improve image compression. Attention modules can help the learned model pay more attention to complex regions. However, many of them are time-consuming, or can only capture local information \cite{zou2022devil}. At the same time, these attention modules are usually placed in both the main and the hyper-prior path of image compression network, which will further introduce large complexity because of a large input size for the main path. 
To overcome that problem, we try to move attention modules to the channel-wise entropy model which has $1/16$ input size compared with that of main path to reduce complexity. Nevertheless, if the above attention modules are directly added to the entropy model, a large number of parameters will be introduced. Therefore, we propose a parameter-efficient swin-transformer-based attention module (SWAtten) with channel squeezing for the channel-wise entropy model. At the same time, to avoid the latency caused by too many slices, we reduce the number of slices from 10 to 5 to achieve the balance between running speed and RD-performance. As Fig. \ref{fig:v} shows, our method can get pleasant results compared with other methods.

\par
The contributions of this paper can be summarized as follows:
\begin{itemize}
    \item We propose a LIC framework with parallel transformer-CNN mixture (TCM) blocks that efficiently incorporate the local modeling ability of CNN and the non-local modeling ability of transformers, while maintaining controllable complexity.
    \item We design a channel-wise auto-regressive entropy model by proposing a parameter-efficient swin-transformer-based attention (SWAtten) module with channel squeezing.
    \item Extensive experiments demonstrate that our approach achieves state-of-the-art (SOTA) performance on three datasets (i.e., Kodak, Tecnick, and CLIC datasets) with different resolutions. The method outperforms VVC (VTM-12.1) by 12.30\%, 13.71\%, 11.85\%  in Bjøntegaard-delta-rate (BD-rate) \cite{bdrate} on Kodak, Tecnick, and CLIC datasets, respectively.
\end{itemize}

\section{Related Work}
\subsection{Learned end-to-end Image Compression}

\subsubsection{CNN-based Models} 
In the past decade, learned image compression has made significant progress and demonstrated impressive performance. Ball\'{e} \etal \cite{balle2016end} firstly proposed an end-to-end learned CNN-based image compression model. Then they proposed a VAE architecture and introduced a hyper-prior to improve image compression in \cite{balle2018variational}. Furthermore, a local context model was utilized to improve the entropy model of image compression in \cite{minnen2018joint}. In addition to that, a causal context was proposed by using global context information in \cite{guo2021causal}. Since the context model is time-consuming, He \etal \cite{he2021checkerboard} designed a checkerboard context model to achieve parallel computing, while Minnen \etal \cite{minnen2020channel} used channel-wise context to accelerate the computing. Apart from improving the entropy model, some works attempt to adopt different types of convolutional neural networks to enhance image compression, such as Cheng \etal \cite{cheng2020learned} who developed a residual network. Chen \etal \cite{chen2022two} introduced octave residual networks into image compression models, and Xie \etal \cite{xie2021enhanced} used invertible neural networks (INNs) to improve performance.
\subsubsection{Transformer-based Models}
With the rapid development of vision transformers, transformers show impressive performance on not only high-level vision tasks, such as, image classification \cite{liu2021swin,touvron2021training}, but also some low-level vision tasks, such as, image restoration \cite{liang2021swinir}, and image denoising \cite{zhang2022practical}. Motivated by those works, some transformer-based LIC models are also proposed recently. Some works \cite{zou2022devil,zhu2022transformerbased} tried to construct a swin-transformer-based LIC model. Qian \etal \cite{qian2021entroformer} used a ViT \cite{dosovitskiy2020image} to help the entropy model capture global context information. Koyuncu \etal \cite{koyuncu2022contextformer} utilized a sliding window to reduce the complexity of ViT in entropy models. Kim \etal \cite{kim2022joint} proposed an Information Transformer to get both global and local dependencies.

\subsection{Attention Modules}
Attention modules try to help the learned models focus on important regions to obtain more details. Many attention modules designed for image compression significantly improve the RD-performance. Liu \etal \cite{liu2019non} firstly introduced a non-local attention module into image compression. Because of the non-local block, this attention module is time-consuming. Therefore, Cheng \etal \cite{cheng2020learned} removed this non-local block and proposed a local attention module to accelerate the computing. Furthermore, Zou \etal \cite{zou2022devil} adopted a window-based attention module to improve image compression. 

\begin{figure*}[h!]
        \centering
        \includegraphics[scale=0.7]{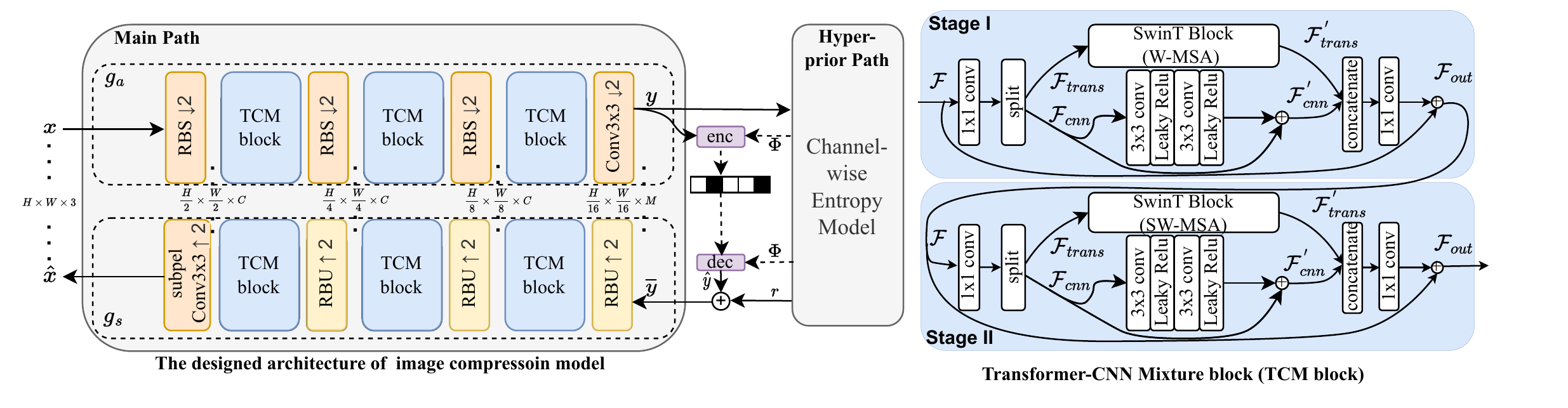}
        \vspace{-0.4cm}
        \caption{The overall framework of our method (left) and the proposed TCM block (right), enc and dec contain the processes of quantizing and range coding. $\uparrow2$ or $\downarrow2$ means that the feature size is enlarged/reduced by two times compared to the previous layer.}
        \vspace{-0.2cm}
        \label{fig:1}
\end{figure*}
\section{Proposed Method}
\subsection{Problem Formulation}
\begin{figure*}[h!]
\vspace{-0.4cm}
        \centering
        \includegraphics[scale=0.75]{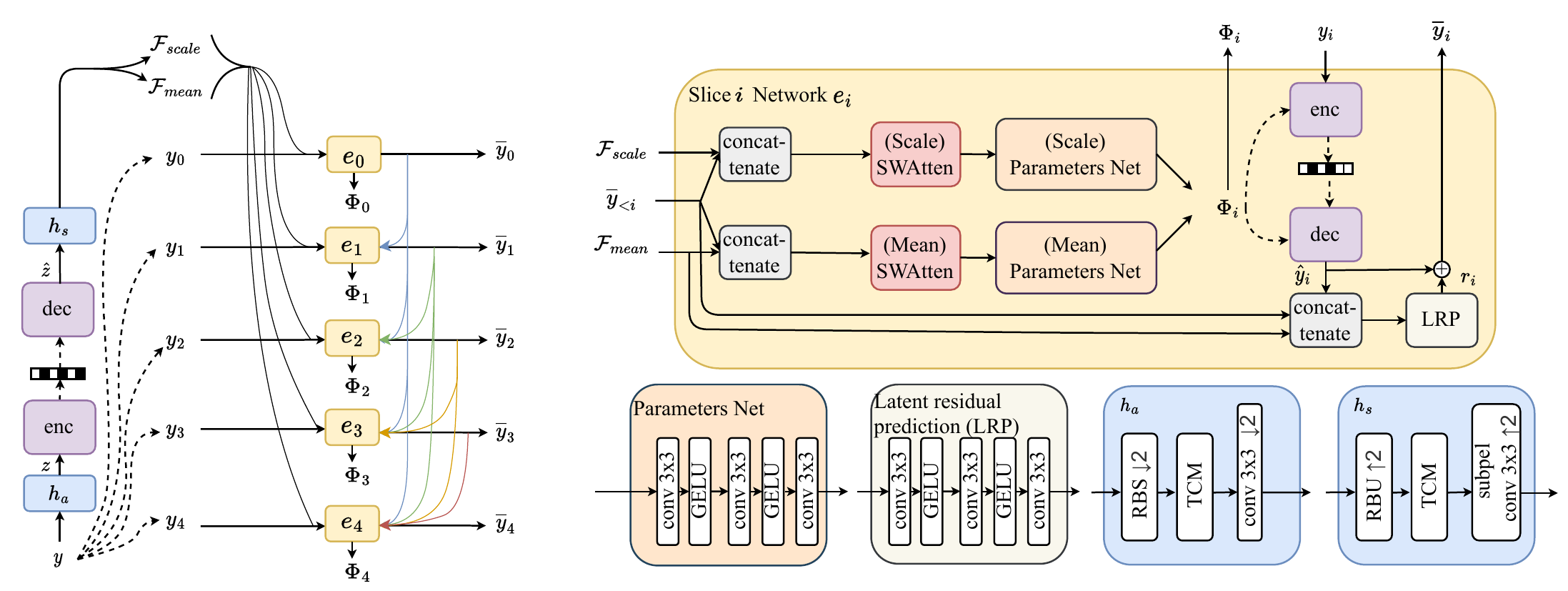}
        \vspace{-0.4cm}
        \caption{The proposed channel-wise entropy model. The encoded slice $\boldsymbol{\overline{y}}_{<i}$ can assist the encoding of the subsequent slice $\boldsymbol{y}_i$.}
        \label{fig:2}
        \vspace{-0.1cm}
\end{figure*}

As Fig.\ref{fig:1} shows, the LIC models with a channel-wise entropy model \cite{minnen2020channel} can be formulated by:
\begin{equation}
\begin{aligned}
&\boldsymbol{y}=g_{a}(\boldsymbol{x} ; \boldsymbol{\phi}) \\
&\hat{\boldsymbol{y}}=Q(\boldsymbol{y}-\boldsymbol{\mu}) + \boldsymbol{\mu}  \\
&\hat{\boldsymbol{x}}=g_{s}(\hat{\boldsymbol{y}} ; \boldsymbol{\theta})
\end{aligned}
\label{quant}
\end{equation}
where $\boldsymbol{x}$ and $\boldsymbol{\hat{x}}$ represent the raw images and decompressed images. By  inputting $\boldsymbol{x}$ to the encoder $g_a$ with learned parameters $\boldsymbol{\phi}$, we can get the latent representation $\boldsymbol{y}$ which is estimated to have a mean $\boldsymbol{\mu}$. To encode it, $\boldsymbol{y}$ is quantized to $\boldsymbol{\hat{y}}$ by quantization operator $Q$. According to previous works \cite{he2022elic, minnen2018joint} and discussion\footnote{\url{https://groups.google.com/g/tensorflow-compression/c/LQtTAo6l26U/m/mxP-VWPdAgAJ}}, we round and encode each $\lceil \boldsymbol{y}-\boldsymbol{\mu}]$ to the bitstream instead of $\lceil \boldsymbol{y}\rfloor$ and restore the coding-symbol $\boldsymbol{\hat{y}}$ as $\lceil \boldsymbol{y}-\boldsymbol{\mu}\rfloor+\boldsymbol{\mu}$, which can further benefit entropy models. Then, we use range coder to encode losslessly $\lceil \boldsymbol{y}-\boldsymbol{\mu}]$ which is modeled as a single Gaussian distribution with the variance $\boldsymbol{\sigma}$ to bitstreams, and transmit it to decoder $g_s$. In this overall pipline, $\boldsymbol{\Phi} = (\boldsymbol{\mu},\boldsymbol{\sigma})$ in this paper is derived by a channel-wise entropy model as Equation \ref{eq2}, Fig. \ref{fig:1} and Fig. \ref{fig:2} show. The entropy model of \cite{minnen2020channel} divides $\boldsymbol{y}$ to $s$ even slices $\{\boldsymbol{y}_0,\boldsymbol{y}_1,...,\boldsymbol{y}_{s-1}\}$ so that the encoded slices can help improve the encoding of subsequent slices, we formulate it as:
\vspace{-0.5cm}

\begin{equation}
\begin{aligned}
\boldsymbol{z}=h_{a}\left(\boldsymbol{y} ; \boldsymbol{\phi}_{h}\right),
 &\quad \hat{\boldsymbol{z}}=Q(\boldsymbol{z}) \\
\boldsymbol{\mathcal{F}}_{mean}, \boldsymbol{\mathcal{F}}_{scale} &= h_s(\hat{\boldsymbol{z}};\boldsymbol{\theta}_h) \\
\boldsymbol{r}_{i}, \boldsymbol{\Phi}_i = e_i(\boldsymbol{\mathcal{F}}_{mean}, &\boldsymbol{\mathcal{F}}_{scale}, \overline{\boldsymbol{y}}_{<i}, \boldsymbol{y}_{i}) ,\ 0<=i<s \\
\overline{\boldsymbol{y}}_{i} &= \boldsymbol{r}_{i} + \hat{\boldsymbol{y}}_{i}
\end{aligned}
\label{eq2}
\end{equation}
where $h_a$ denotes the hyper-prior encoder with parameters $\boldsymbol{\phi}_{h}$. It is used to get side information $\boldsymbol{z}$ to capture spatial dependencies among the elements of $\boldsymbol{y}$. A factorized density model $\boldsymbol{\psi}$ is used to encode quantized $\boldsymbol{\hat{z}}$ as $p_{\hat{\boldsymbol{z}} \mid \boldsymbol{\psi}}(\hat{\boldsymbol{z}} \mid \boldsymbol{\psi})=\prod_{j}\left(p_{z_{j} \mid \boldsymbol{\psi}}(\boldsymbol{\psi}) * \mathcal{U}\left(-\frac{1}{2}, \frac{1}{2}\right)\right)\left({\hat{z}}_{j}\right)$ where $j$ specifies the position of each element or each signal. $\hat{\boldsymbol{z}}$ is then fed to the hyper-prior decoder $h_s$ with parameters $\boldsymbol{\theta}_h$ for decoding to obtain two latent features $\boldsymbol{\mathcal{F}}_{mean}$, $\boldsymbol{\mathcal{F}}_{scale}$ which are used to be input to the following each slice network $e_i$.
After that, each slice $\boldsymbol{y}_i$ is sequentially processed to get $\boldsymbol{\overline{y}_i}$. During this process, encoded slices $\boldsymbol{\overline{y}}_{<i}=\{\boldsymbol{\overline{y}}_0, \boldsymbol{\overline{y}}_1, ..., \boldsymbol{\overline{y}}_{i-2}, \boldsymbol{\overline{y}}_{i-1}\}$ and current slice $\boldsymbol{y}_i$ are input to the slice network $e_i$ to get the estimated distribution parameters $\boldsymbol{\Phi}_i=(\boldsymbol{\mu}_i,\boldsymbol{\sigma}_i)$ to help generate bit-streams. Therefore, we can assume $p_{\hat{\boldsymbol{y}} \mid \hat{\boldsymbol{z}}}(\hat{\boldsymbol{y}} \mid \hat{\boldsymbol{z}}) \sim \mathcal{N}\left(\boldsymbol{\mu}, \boldsymbol{\sigma}^{2}\right)$. At the same time, the residual $\boldsymbol{r}_i$ is used to reduce the quantization errors ($\boldsymbol{y}-\boldsymbol{\hat{y}}$) which is introduced by quantization. Therefore, $\boldsymbol{\overline{y}}$ with less error is entered into the decoder $g_s$ with learned parameters $\boldsymbol{\theta}$, instead of $\boldsymbol{\hat{y}}$ in Equation \ref{quant}. At last, we can get the decompressed image $\boldsymbol{\hat{x}}$. Fig. \ref{fig:2} illustrates the detailed process of this channel-wise entropy model clearly.

\par 
In order to train the overall learned image compression model, we consider the problem as a Lagrangian multiplier-based rate-distortion optimization. The loss is defined as:
\begin{equation}
\begin{aligned}
\mathcal{L}=& \mathcal{R}(\hat{\boldsymbol{y}})+\mathcal{R}(\hat{\boldsymbol{z}})+\lambda \cdot \mathcal{D}(\boldsymbol{x}, \hat{\boldsymbol{x}}) \\
=& \mathbb{E}\left[-\log _{2}\left(p_{\hat{\boldsymbol{y}} \mid \hat{\boldsymbol{z}}}(\hat{\boldsymbol{y}} \mid \hat{\boldsymbol{z}})\right)\right]+\mathbb{E}\left[-\log _{2}\left(p_{\hat{\boldsymbol{z}} \mid \boldsymbol{\psi}}(\hat{\boldsymbol{z}} \mid \boldsymbol{\psi})\right)\right] \\
&+\lambda \cdot \mathcal{D}(\boldsymbol{x}, \hat{\boldsymbol{x}})
\end{aligned}
\label{equ:RD}
\end{equation}
where $\lambda$ controls the rate-distortion tradeoff. Different $\lambda$ values are corresponding to different bit rates. $\mathcal{D}(\boldsymbol{x}, \hat{\boldsymbol{x}})$ denotes the distortion term which is calculated by Mean squared error (MSE) loss. $\mathcal{R}(\boldsymbol{\hat{y}}), \mathcal{R}(\boldsymbol{\hat{z}})$ denote the bit rates of latents $\boldsymbol{\hat{y}}$ and $\boldsymbol{\hat{z}}$.

\begin{figure}[h!]
        \centering
        \includegraphics[scale=0.47]{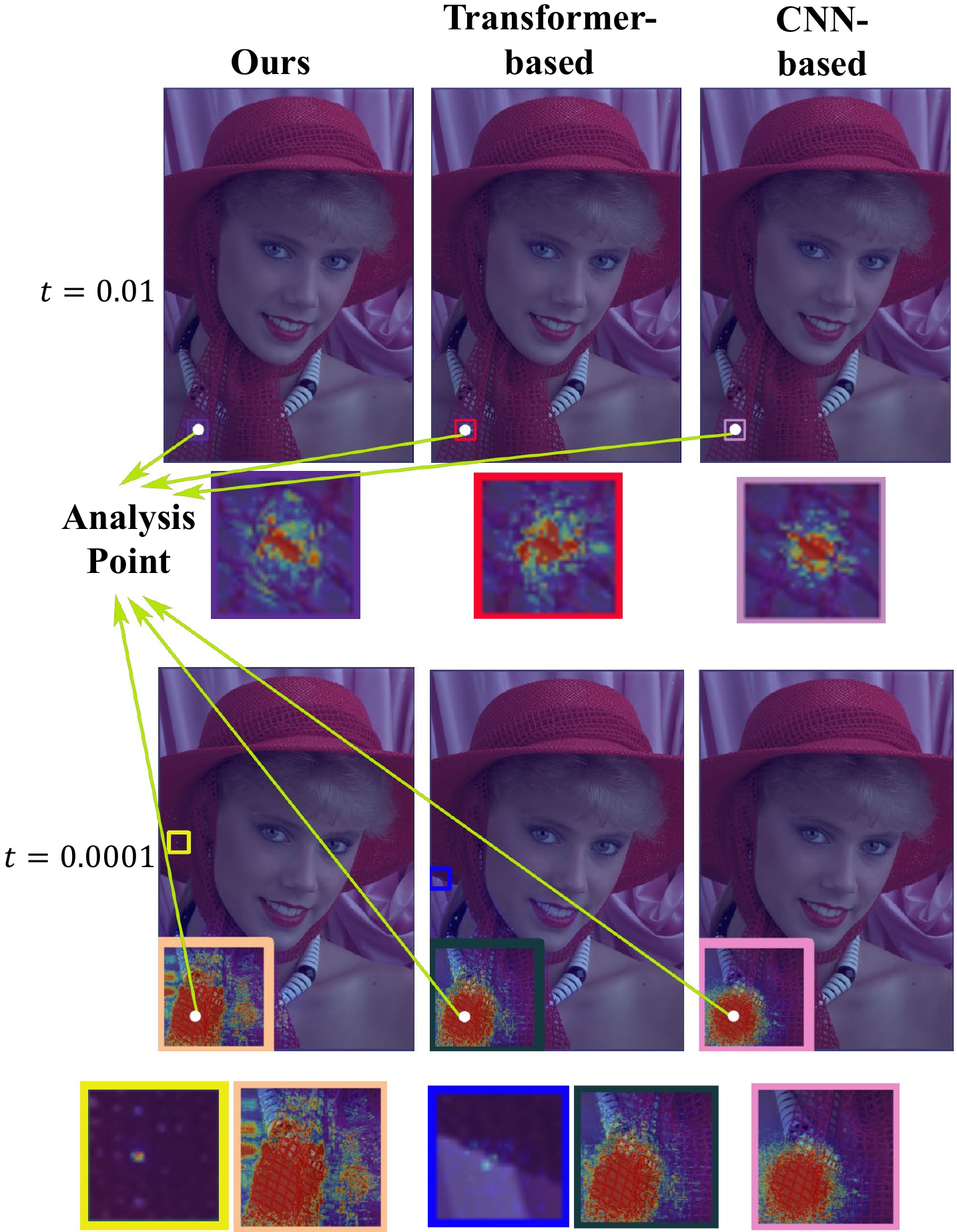}
        \caption{The effective receptive fields (ERF) calculated by different methods and clipped by different thresholds $t$. The two rows correspond to different thresholds while the three columns correspond to different methods including our proposal, transformer-based \cite{zou2022devil}, and CNN-based \cite{minnen2020channel} method.}
        \vspace{-0.5cm}
        \label{fig:22}
\end{figure}

\begin{figure*}[h!]
\centering
        \includegraphics[scale=0.5]{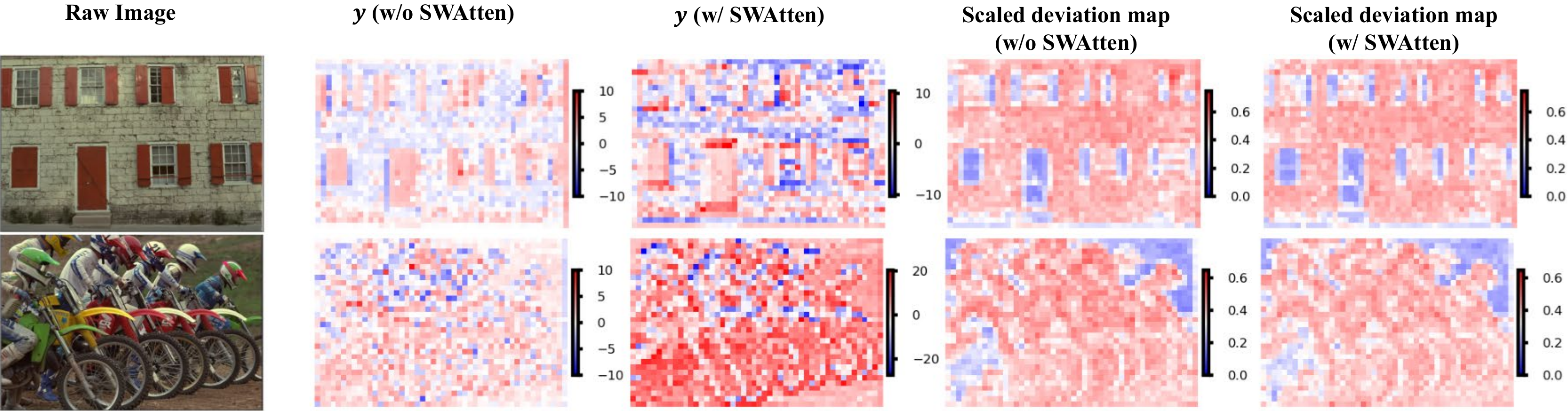}
	\caption{The scaled deviation map and the channel with the largest entropy of latent representations $y$, from the model without/with SWAtten. For $kodim01$ (top), the [PSNR@bitrates@$\epsilon_s$] of the model without and with SWAtten are [29.61dB@0.503bpp@0.451] and [29.95dB@0.501bpp@0.389]. For $kodim05$ (bottom), they are [30.07dB@0.537bpp@0.422] and [31.03dB@0.534bpp@0.365].}
	\vspace{-0.5cm}
	\label{fig:5}
\end{figure*}
\subsection{Transformer-CNN Mixture Blocks}
Image compression models based on CNN \cite{cheng2020learned} have achieved excellent RD-performance. Besides, with the rapid development of vision transformers, some methods based on vision transformers \cite{zhu2022transformerbased} are also proposed and outperform CNN-based methods because transformers can capture non-local information. However, according to previous works \cite{zou2022devil, he2021checkerboard}, even though non-local information can improve image compression, local information still has a significant impact on the performance of image compression. CNN can pay more attention to local patterns while transformers have the ability of non-local information. Therefore, we try to incorporate residual networks and Swin-Transformer (SwinT) blocks \cite{liu2021swin} to utilize both advantages of these two kinds of models. There are two challenges in this combination, the first is how to fuse two different features effectively, and the other is how to reduce the required complexity. 
\par
Here, an efficient parallel transformer-CNN mixture (TCM) block is proposed as Fig. \ref{fig:1} shows. We assume the input tensor as $\boldsymbol{\boldsymbol{\mathcal{F}}}$ with size $C\times H_{\boldsymbol{\mathcal{F}}}\times W_{\boldsymbol{\mathcal{F}}}$. It is firstly input into a $1\times 1$ convolutional layer whose output channels number is also $C$. Then we evenly split the tensor to two tensor $\boldsymbol{\boldsymbol{\mathcal{F}}}_{cnn}$ and $\boldsymbol{\boldsymbol{\mathcal{F}}}_{trans}$ with size $\frac{C}{2}\times H_{\boldsymbol{\mathcal{F}}}\times W_{\boldsymbol{\mathcal{F}}}$. 
This operation has two benefits. Firstly, it reduces the number of feature channels fed to subsequent CNN and transformers to decrease the model complexity. Secondly, local and non-local features can be processed independently and in parallel, which facilitates better feature extraction.
After that, the tensor $\boldsymbol{\mathcal{F}}_{cnn}$ is sent to the residual network (Res) to get $\boldsymbol{\mathcal{F}}^{'}_{cnn}$, while $\boldsymbol{\mathcal{F}}_{trans}$ is sent to the SwinT Blocks to get $\boldsymbol{\mathcal{F}}^{'}_{trans}$. Then, we concatenate $\boldsymbol{\mathcal{F}}^{'}_{trans}$ and $\boldsymbol{\mathcal{F}}^{'}_{cnn}$ to get a tensor with size $C\times H_{\boldsymbol{\mathcal{F}}}\times W_{\boldsymbol{\mathcal{F}}}$. And then, the concatenated tensor is input to another $1\times 1$ convolutional layer to fuse local and non-local features. At last, the skip connection between $\boldsymbol{\mathcal{F}}$ and the output is built to get $\boldsymbol{\mathcal{F}}_{out}$. In order to combine the residual network with the Swin-Transformer more effectively, we divide TCM into two stages with similar processes. In the transformer block of stage I, we use window-based multi-head self-attention (W-MSA). In stage II, we use shifted window-based multi-head self-attention (SW-MSA) in the transformer block. The benefit of this is that the residual networks are inserted into the common two consecutive Swin-transformer blocks, which can be more effective for feature fusion. Both two stages can be formulated as follows:
\begin{equation}
    \begin{aligned}
    &\boldsymbol{\mathcal{F}}_{cnn}, \boldsymbol{\mathcal{F}}_{trans} = Split(Conv1\times1(\boldsymbol{\mathcal{F}})) \\
    & \boldsymbol{\mathcal{F}^{'}}_{cnn}, \boldsymbol{\mathcal{F}^{'}}_{trans} = Res(\boldsymbol{\mathcal{F}}_{cnn}), SwinT(\boldsymbol{\mathcal{F}}_{trans}) \\
    &\boldsymbol{\mathcal{F}}_{out}=\boldsymbol{\mathcal{F}}+  Conv1\times1(Cat(\boldsymbol{\mathcal{F}^{'}}_{cnn}, \boldsymbol{\mathcal{F}^{'}}_{trans}))
    \end{aligned}
\end{equation}
\par
Based on the proposed TCM Block, the main path ($g_a$ and $g_s$) is designed as Fig. \ref{fig:1} shows. The Residual Block with stride (RBS), Residual Block Upsampling (RBU) and subpixel conv3x3 are proposed by \cite{cheng2020learned}. The detailed architectures of these three modules are reported in Supplementary. 
In our framework, except for the last layer of $g_a$ and $g_s$, we attach a TCM block after each RBS/RBU to obtain non-local and local information. 
Also, we add TCM blocks into the hyper-prior path by re-designing the hyper-prior encoder $h_a$ and hyper-prior decoder $h_s$ as Fig. \ref{fig:2} shows.
\par
To explore how the TCM block aggregates local and non-local information, we present effective receptive fields (ERF) \cite{luo2016understanding} of our model. Besides, the effective receptive fields of the transformer-based model \cite{zou2022devil} and the CNN-based model \cite{minnen2020channel} are used to make comparisons. ERF is defined as absolute gradients of a pixel in the output (i.e., $|\frac{d\hat{x}_p}{d\boldsymbol{x}}|$). Here, we calculate gradients of the analysis point $p=(70,700)$ of $kodim04$ in Kodak dataset. In order to estimate the importance of information at different distances, we clip gradients with two thresholds $t$ (i.e., 0.01, 0.0001) to visualize them. Note that the clip operation means we reduce the gradient values larger than the threshold to the threshold, and the gradient values smaller than the threshold remain unchanged. This can help our visualization has higher visibility precision. The visualization is shown in Fig. \ref{fig:22}. As we can see, when $t=0.01$, the red regions (high gradient values) of our TCM-based model and the CNN-based model are smaller than that of the transformer-based model. This suggests that our model pays more attention to neighbor regions, and has a similar local modeling ability of CNN. Meanwhile, when $t=0.0001$, the ERF show our model and the transformer-based model can capture information at a long distance (As shown by the two small boxes in yellow and dark blue in Fig. \ref{fig:22}, there is still a gradient return at a distance far from point $p$). This means that our model also has the long-distance modeling ability of transformers. It is also worth noting that at $t=0.0001$, the CNN-based model exhibits a circular ERF, while the ERF of our model exhibit a shape closer to the context (a long strip shape like the background scarf). This shows that our model has better modeling ability compared to the CNN-based model.

\subsection{Proposed Entropy Model}
Motivated from \cite{minnen2020channel, he2022elic}, we propose a channel-wise auto-regressive entropy
model with a parameter-efficient swin-transformer-based attention module (SWAtten) by using channel squeezing. The framework is shown in Fig. \ref{fig:2}.
\subsubsection{SWAtten Module}
\begin{figure}[h!]
        \centering
        \includegraphics[scale=0.5]{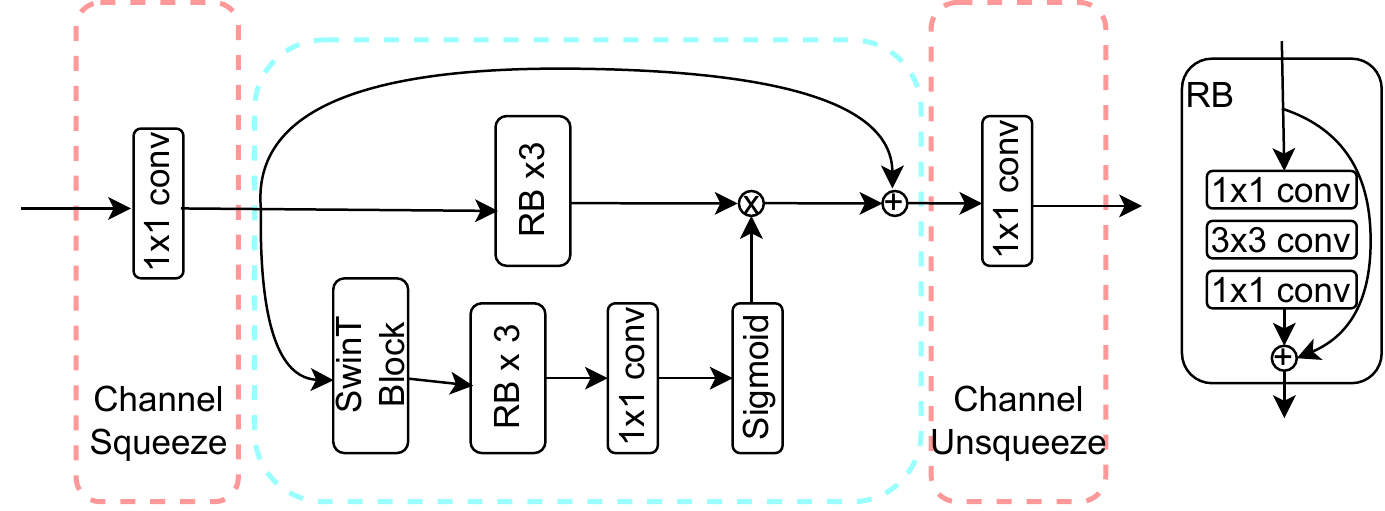}
        \caption{The proposed SWAtten module.}
        \label{fig:23}
        \vspace{-2mm}
\end{figure}

Past works on attention have demonstrated their effectiveness for image compression. However, many of them are time-consuming, or only have the ability to capture local information. Different from these modules which are placed on both the main path ($g_a$ and $g_s$) and hyper-prior path ($h_a$ and $h_s$) of image compression, we design an attention module for the entropy model which has $1/16$ input size compared to the main path and can reduce much complexity. The designed parameter-efficient swin-Transformer based Attention (SWAtten) module is shown in Fig. \ref{fig:23}. The swin-transformer block which can capture non-local information is added into the architecture, while the other residual blocks (RB) can get local information. Since the number of channels of features inputted to $e_i$ accumulates with the increase of slice index $i$, the input channels of $e_i$ can be expressed as:
\begin{equation}
    M + i \times (M//s)
\end{equation}
Where $M$ is the number of channels of the latent variable $y$. The input channel number of $e_9$ can reach 608 when $s$ is set as 10 and $M$ is set as 320, which causes the model to be parameter-consuming.
To achieve the balance between complexity and RD-performance, the total number of the slices $s$ is reduced to 5 from 10 which is the common setting in \cite{minnen2020channel}. At the same time, a channel squeeze operation is used to squeeze the input channels. In this paper, we squeeze the input channels of all slices to 128, i.e., let the output channel of the first $1\times 1$ convolutional layer be 128. At last, an unsqueeze operation is utilized to unsqueeze the channels of output to the original number, i.e., let the output channel of the last 1x1 convolutional layer be $M + i \times (M/s)$.

\par
It should be noted that although both SWAtten and TCM are composed of transformers and CNNs, there are some differences between them. Firstly, according to \cite{luo2022}, the hyper-prior path of the image compression network contains numerous redundant parameters, making it possible to use channel squeezing to greatly reduce the parameters without compromising performance. But the main path is sensitive to parameters, which means we cannot use such operations. Secondly, CNN in SWAtten is used not only for extracting local features but also for extracting attention maps, which differs from TCM. Lastly, the receptive field of the entropy model in SWAtten is large enough, eliminating the need for a two-stage fusion framework like TCM.

\par
According to \cite{xie2021enhanced}, the deviation between $\boldsymbol{\hat{y}}$ and $\boldsymbol{{y}}$ with size $M\times H_y\times W_y$ can be used to analyze the information loss in the process of compression, we can formulate the mean absolute pixel deviation $\epsilon$ as:
\vspace{-0.5cm}

\begin{equation}
\begin{aligned}
&\epsilon =\sum_{h=1}^{H_y} \sum_{w=1}^{W_y} \sum_{m=1}^{M} |{\hat{y}}_{h,w,m} - {{y}}_{h,w,m}| \\
& =\sum_{h=1}^{H_y} \sum_{w=1}^{W_y} \sum_{m=1}^{M} |Q({{y}}_{h,w,m}-{\mu}_{h,w,m}) - ({y}_{h,w,m}-{\mu}_{h,w,m})|
\end{aligned}
\end{equation}
\par
To compare the deviation among different models, it is unfair to directly compare the values $\epsilon$ because the $\boldsymbol{\hat{y}}$ and $\boldsymbol{y}$ in different models have different ranges. Since the deviation is relative to $\boldsymbol{y}$, a scaling factor $\gamma=\sum_{h=1}^{H_y} \sum_{w=1}^{W_y} \sum_{m=1}^{M}|{y}_{h,w,m}|$ is introduced to define a scaled mean absolute pixel deviation $\epsilon_s$ to better evaluate the information loss. $\epsilon_s$ can be formulated as:
\begin{equation}
    \epsilon_s = \epsilon/\gamma
\end{equation}
\par 
Fig. \ref{fig:5} shows the scaled deviation map and the channel with the largest entropy of $\boldsymbol{y}$ of $kodim01$ and $kodim05$ in Kodak dataset by using the model with SWAtten or not. 
Note that the deviation map is in shape $(H_y, W_y)$ and each pixel is the mean of the absolute deviation along the channel dimension after scaling with $\gamma$.
The $\epsilon_s$ for $kodim01$ and $kodim05$ can be reduced from 0.451 and 0.422 by using the model without SWAtten to 0.389 and 0.365 by using the model with SWAtten. It suggests that the model with SWAtten can have less information loss and get a higher quality decompressed image. 

\begin{figure*}[h!]
\captionsetup[subfigure]{labelformat=empty}
  \centering
 \begin{subfigure}{.46\linewidth}
    \centering\includegraphics[width=1.\linewidth]{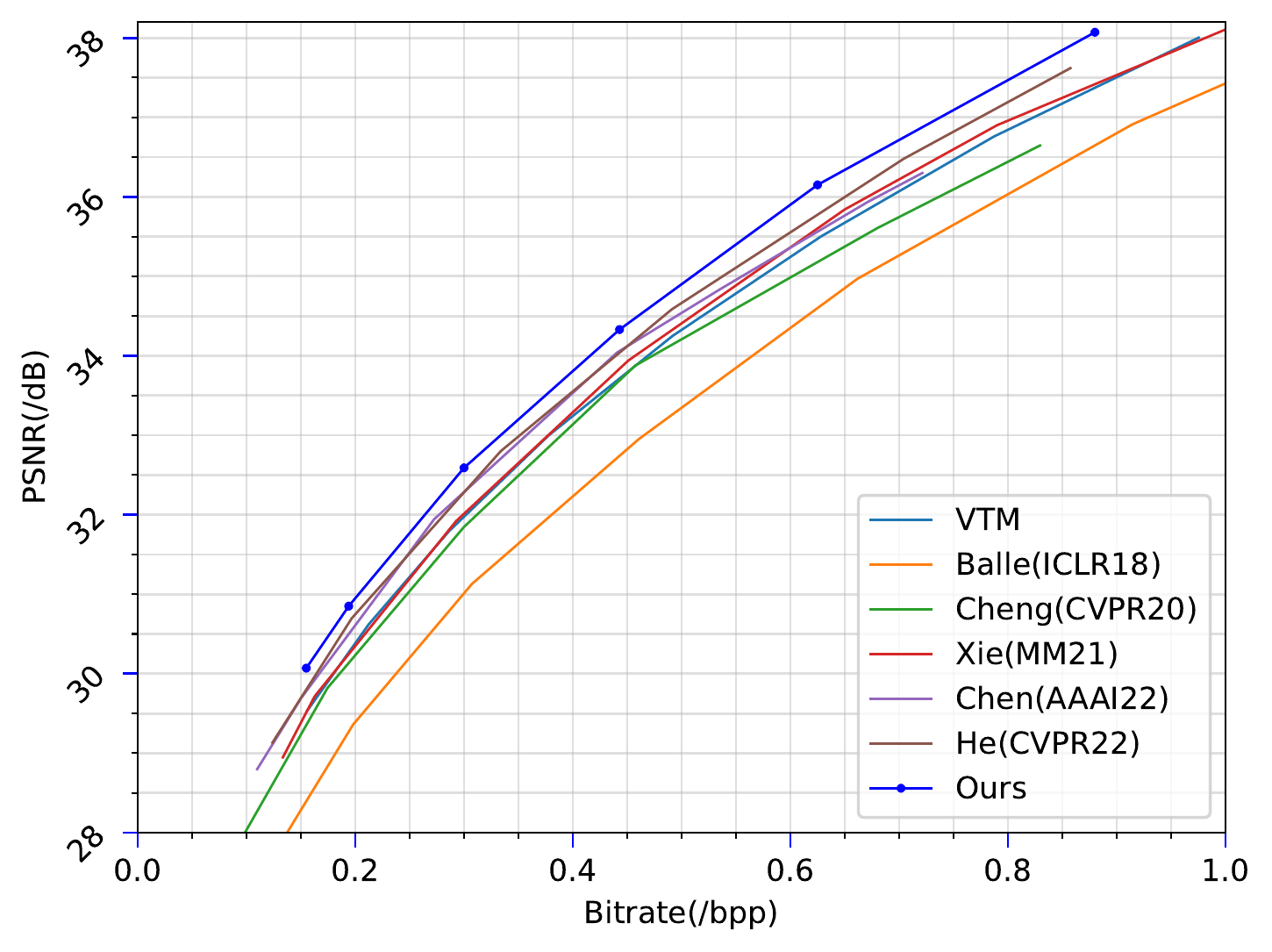}
  \end{subfigure}
  \begin{subfigure}{.46\linewidth}
    \centering\vspace{-0.5cm}\includegraphics[width=1.1\linewidth]{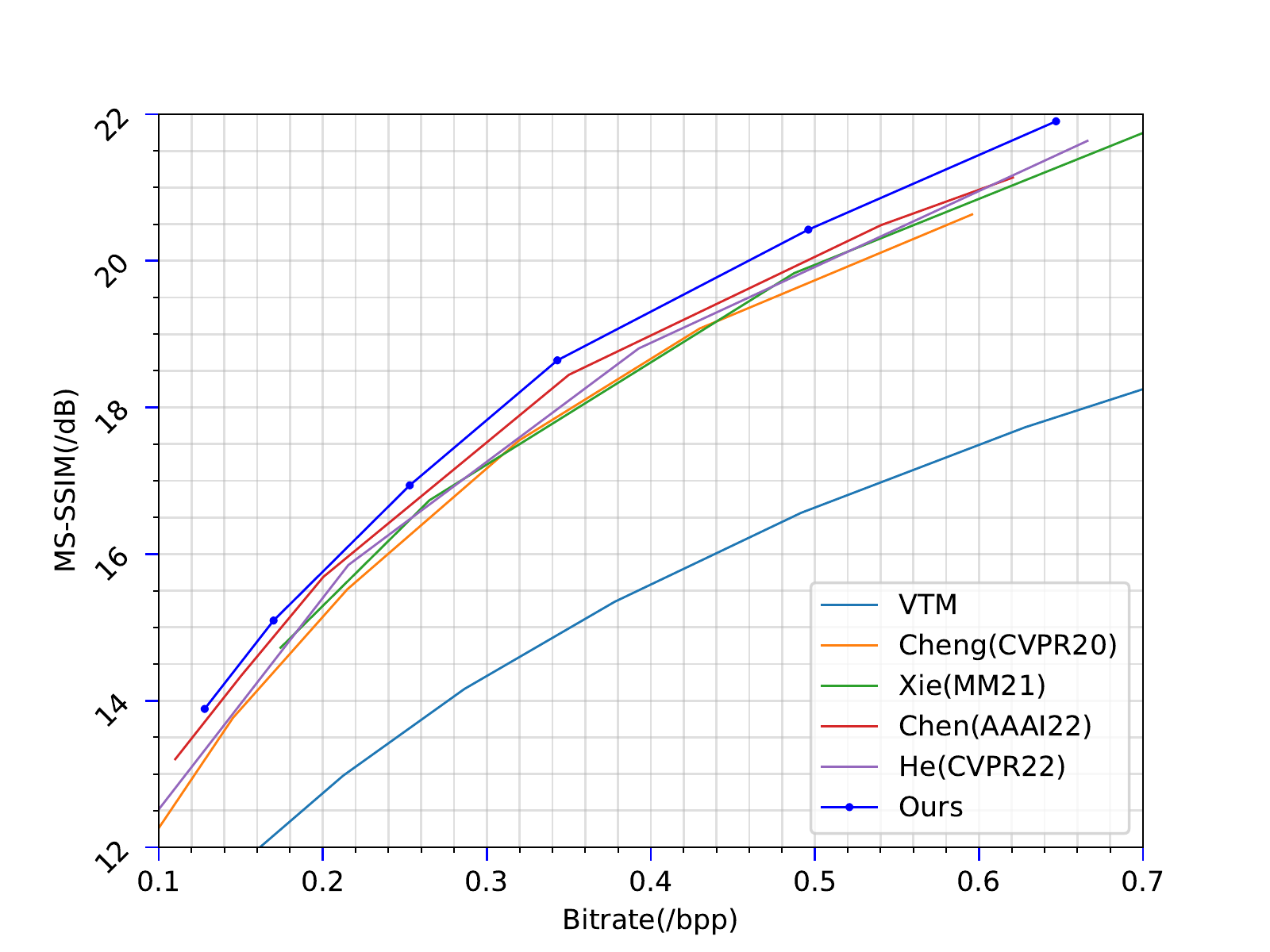}
  \end{subfigure}
  \vspace{-0.4cm}
  \caption{Performance evaluation on the Kodak dataset.}
  \label{fig:kodak}
\end{figure*}
\begin{figure*}[htbp]
\vspace{-0.4cm}
\centering
\begin{minipage}[t]{0.47\textwidth}
\centering
\includegraphics[width=8.2cm]{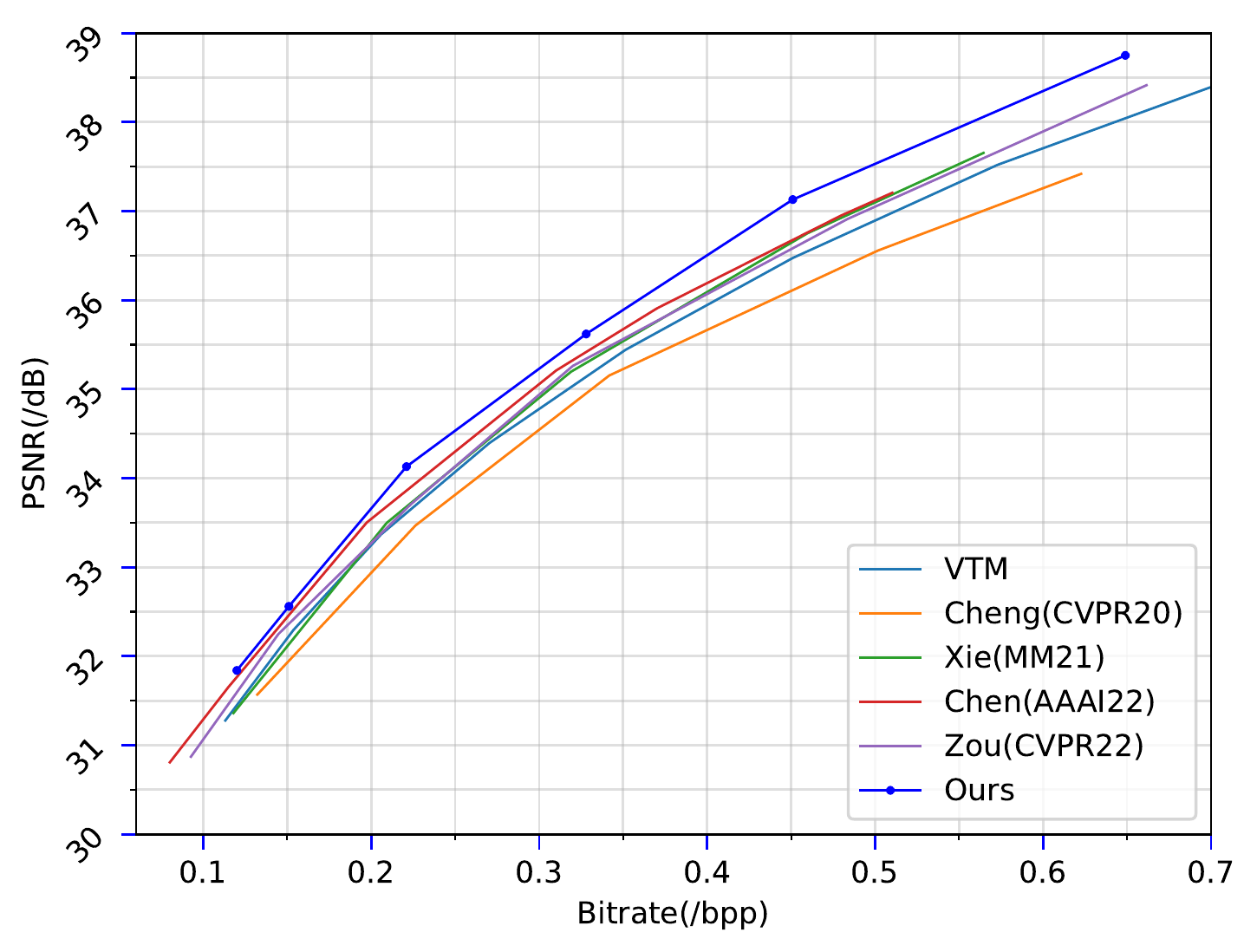}
\caption{Performance evaluation on the CLIC Professional Validation dataset.}
\label{fig:6}
\end{minipage}
\begin{minipage}[t]{0.47\textwidth}
\centering
\includegraphics[width=8.2cm]{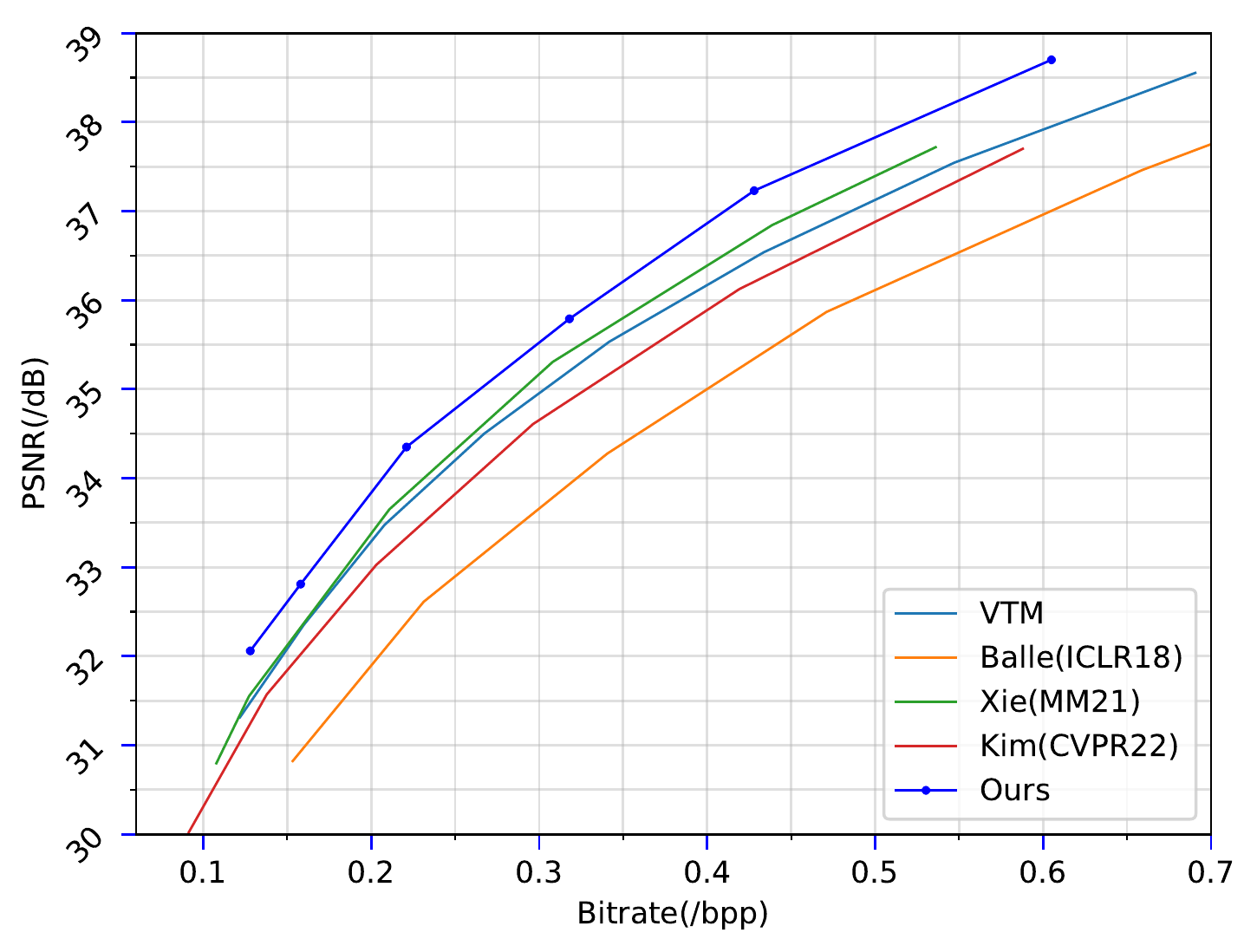}
\caption{Performance evaluation on the Tecnick dataset.}
\label{fig:7}
\end{minipage}
\vspace{-6mm}
\end{figure*}

\section{Experiments}
\subsection{Experimental Setup}
\subsubsection{Training Details}
For training, we randomly choose 300k images of size larger than $256\times256$ from ImageNet \cite{deng2009large}, and randomly crop them with the size of $256\times256$ during the training process. We adopt Adam \cite{kingma2014adam} with a batch size 8 to optimize the network. The initial learning rate is set as $1\times 10^{-4}$. After 1.8M steps, the learning rate is reduced to $1\times 10^{-5}$ for the last 0.2M steps. 
\par 
The model is optimized by RD-formula as Equation \ref{equ:RD}. Two kinds of quality metrics, i.e., mean square error (MSE) and MS-SSIM, are used to represent the distortion $\mathcal{D}$. When the model is optimized by MSE, the $\lambda$ belongs to $\{0.0025,0.0035,0.0067,0.0130,0.0250,0.0500\}$. When the model is optimized by MS-SSIM, the $\lambda$ belongs to $\{3,5,8,16,36,64\}$. 
\par 
For swin-transformer blocks, window sizes are set as 8 in the main path ($g_a$ and $g_s$), and 4 in the hyper-prior path ($h_a$ and $h_s$). The channel number $M$ of the latent $\boldsymbol{y}$ is set as 320, while that of $\boldsymbol{z}$ is set as 192, respectively. Other hyper-parameters in the entropy model follow the setting in \cite{minnen2020channel}. We use RTX 3090 and Intel i9-10900K to complete the following experiments.
\subsubsection{Evaluation}
We test our method on three datasets, i.e., Kodak image set \cite{kodak} with the image size of $768\times 512$, old Tecnick test set\footnote{\url{https://sourceforge.net/projects/testimages/files/OLD/}}\cite{asuni2014testimages} with the image size of $1200\times 1200$, CLIC professional validation dataset\footnote{\url{http://clic.compression.cc/2021/tasks/index.html}}\cite{clic} with 2k resolution. Both PSNR and MS-SSIM are used to measure the distortion, while bits per pixel (bpp) are used to evaluate bitrates.

\subsubsection{Definition of Various Models}
In the experiments, to explore the performance of our model with different complexities, we tested three different models (small, medium and large by setting different channel number $C=128, 192, 256$ in the middle layers). The location of $C$ is shown in Fig. \ref{fig:1}. The number of slices $s$ of the entropy model for all models with attention modules is reduced to 5 from 10 which is a common setting in \cite{minnen2020channel}. More details are reported in Supplementary.

\subsection{Rate-Distortion Performance}
We compare our large model with state-of-the-arts (SOTA) learned end-to-end image compression algorithms, including \cite{balle2018variational}, \cite{cheng2020learned}, \cite{xie2021enhanced}, \cite{chen2022two}, \cite{kim2022joint}, \cite{he2022elic}, and \cite{zou2022devil}. The classical image compression codec, VVC \cite{vvc} is also tested by using VTM12.1.
The rate-distortion performance on Kodak dataset is shown in Fig. \ref{fig:kodak}. Both PSNR and MS-SSIM are tested on Kodak to demonstrate the robustness of our method. Here, we convert MS-SSIM to $-10\text{log}_{10}(1-\text{MS-SSIM})$ for clearer comparison. As we can see, at the same bitrate, we can improve up to about 0.4dB PSNR and 0.5dB MS-SSIM compared with SOTA methods. The results of CLIC dataset and Tecnick dataset are shown in Fig. \ref{fig:6} and Fig. \ref{fig:7}, respectively. We also achieve similar good results on these two datasets. These results suggest that our method is robust and can achieve SOTA performance based on all of the three datasets with different resolutions.
To get quantitative results, we present the BD-rate \cite{bdrate} computed from PSNR-BPP curves as the quantitative metric. The anchor RD-performance is set as the results of VVC on different datasets (BD-rate=0\%). Our method outperforms VVC (VTM-12.1) by 12.30\%, 13.71\%, 11.85\% in BD-rate on Kodak, Tecnick, and CLIC datasets, respectively. 
Table \ref{tab:infer} shows partial results on Kodak. 
More comparisons are reported in Supplementary.

\subsection{Ablation Studies}
\subsubsection{Comparison with Transformer-only/CNN-only based Models}
In order to show the effectiveness of our proposed Transformer-CNN Mixture (TCM) blocks, we compare our medium model without SWAtten modules to the Transformer-only based model and CNN-only based model in \cite{zhu2022transformerbased}. The results are shown in Fig. \ref{fig:9}. ``Conv\_ChARM" and ``SwinT\_ChARM" are a CNN-only based model and a Transformer-only based model, respectively. They have similar architectures to our methods without SWAtten modules. The difference is that ``SwinT\_ChARM" uses Swin Transformer block, ``Conv\_ChARM" uses convolutional neural networks, and we use the proposed TCM block. By using the advantages both of transformer and CNN, the results show that our method surpasses the Transformer-only based model and CNN-only based model. 

\subsubsection{SWAtten Module}
\begin{figure}[htb]
\begin{subfigure}[ht]{0.32\linewidth}
	    \centering
		\includegraphics[scale=0.55]{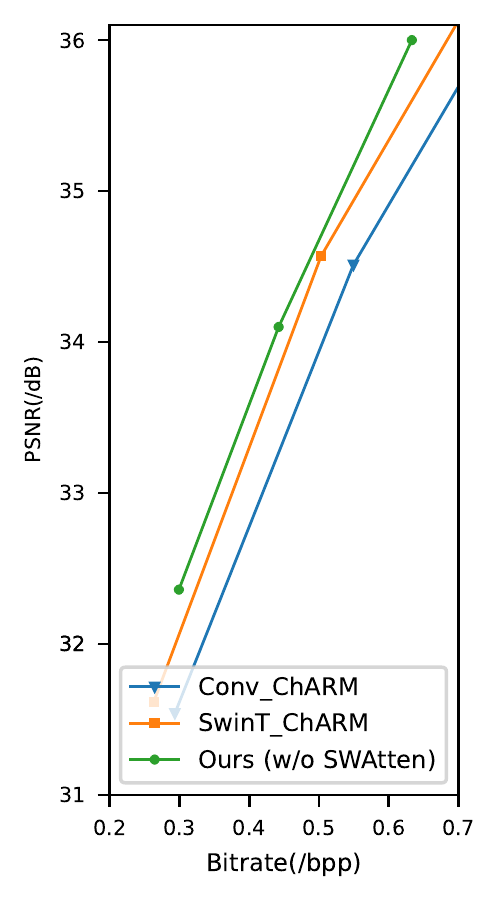}
		\caption{}
		\label{fig:9}
	\end{subfigure}
	\begin{subfigure}[ht]{0.32\linewidth}
	    \centering
		\includegraphics[scale=0.55]{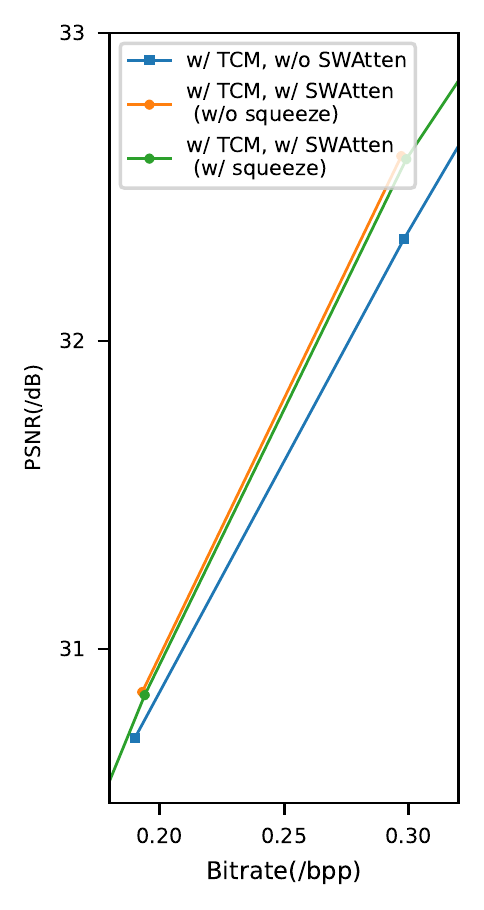}
		\caption{}
		\label{ablation}
	\end{subfigure}
	\begin{subfigure}[ht]{0.32\linewidth}
	    \centering
		\includegraphics[scale=0.55]{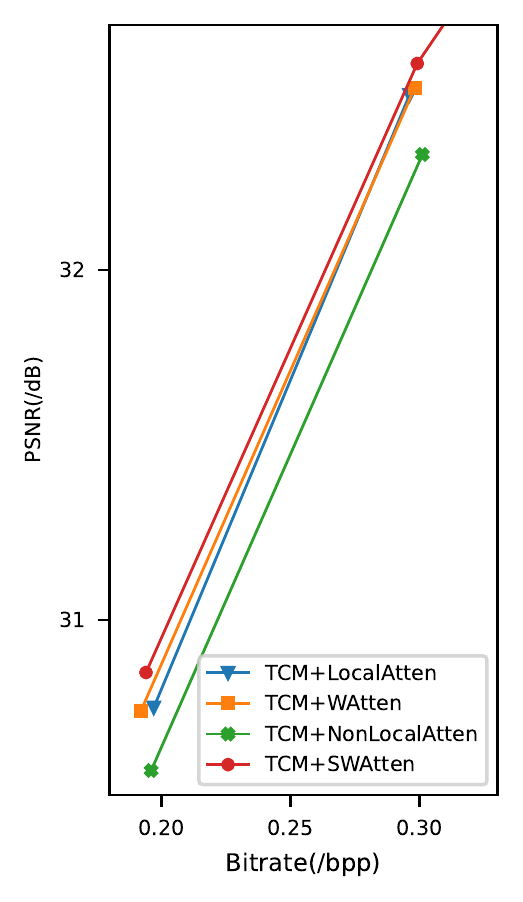}
		\caption{}
		\label{attenex}
	\end{subfigure}
	\caption{Experiments on Kodak dataset. (a) Comparison with Transformer-only/CNN-only based
Models. (b) The ablation studies on SWAtten module (``w/o squeeze" represents that we don't add the channel squeeze/unsqueeze operation to SWAtten). (c) The RD-performance of models using different attention modules.}
\vspace{-0.3cm}
	\label{fig:8}
\end{figure}

In Fig. \ref{ablation}, we compare the cases using SWAtten modules or not. It can be observed that SWAtten modules bring a significant gain in RD-performance. Meanwhile, by using the channel squeeze operation in SWAtten modules, we can get a comparable performance compared with the situation without using that operation while saving many parameters. Section \ref{com} shows more information on the parameters efficiency gain of this operation. 

\subsection{Various Attention Modules}
In Fig. \ref{attenex}, we compared our proposed SWAtten with the previous attention modules, including the non-local attention (NonlocalAtten) module \cite{liu2019non}, the local attention (LocalAtten) module \cite{cheng2020learned}, and the window-based attention (WAtten) module \cite{zou2022devil}. Compared with these different attention modules, SWAtten gets the best RD-performance because it can capture non-local information while also paying enough attention to local information.

\subsection{Complexity and Qualitative Results}
\label{com}
We test the complexity and qualitative results of different methods based on Kodak. Two other SOTA works \cite{xie2021enhanced, zhu2022transformerbased}, are also tested as Table \ref{tab:infer} shows. The results of our method suggest that the efficiency and RD-performance of our method can outperform both of these two methods. Meanwhile, after using channel squeeze in SWAtten, we can save a lot of parameters and FLOPs, while we can get a comparable BD-rate. It also should be noted that all of our small, medium and large models can achieve SOTA RD-performance. Meanwhile, the performance can further improve as the complexity increases, which shows that our model has a lot of potentials.
\begin{table}[htb]
\caption{The RD-performance and complexity of learned image compression models based on Kodak using GPU (RTX 3090). A lower BD-rate indicates higher RD-performance. }
\setlength{\tabcolsep}{0.5mm}
\small
\begin{tabular}{c|c|c|c|c|c}

\hline
\textbf{}                                                                   \textbf{Methods} & \textbf{\begin{tabular}[c]{@{}c@{}}Encoding\\ Time(/ms)\end{tabular}} & \textbf{\begin{tabular}[c]{@{}c@{}}Decoding\\ Time(/ms)\end{tabular}} & \textbf{\begin{tabular}[c]{@{}c@{}}Para-\\ meters(/M)\end{tabular}} & \textbf{\begin{tabular}[c]{@{}c@{}}FLOPs\\ (/G)\end{tabular}} & \textbf{\begin{tabular}[c]{@{}c@{}}BD-\\ rate\end{tabular}} \\ \hline \hline
\textbf{Xie \etal \cite{xie2021enhanced}}                                                   & 2346                                                    & 5212                                                    & 47.55                                                  & 408.21         & -1.65            \\ \hline
\textbf{\begin{tabular}[c]{@{}c@{}}SwinT\\ \_ChARM \cite{zhu2022transformerbased}\end{tabular}} & 132                                                     & 84                                                      & 60.55                                                  & 230.37         & -4.02            \\ \hline

\textbf{\begin{tabular}[c]{@{}c@{}}Ours (Large,\\ w/o squeeze)\end{tabular}} & 151                                                     & 141                                                      & 160.45                                                  & 830.90         & -12.54            \\ \hline

\textbf{Ours(Large)}                                                         & 150                                                     & 140                                                     & 75.89                                                 & 700.96         & -12.30           \\ \hline
\textbf{Ours(Medium)}                                                        & 130                                                     & 122                                                     & 58.72                                                 & 415.20         & -9.65            \\ \hline
\textbf{Ours(Small)}                                                         & 109                                                     & 102                                                     & 44.96                                                 & 211.54         & -7.39            \\ \hline
\end{tabular}
\label{tab:infer}
\vspace{-3mm}
\end{table}

\subsection{Visualization}
Fig. \ref{fig:v} shows the visualization example of decompressed images ($kodim23$) from Kodak dataset by our methods and the classical compression standards WebP, VVC (VTM 12.1) \cite{vvc}. In some complex texture parts, our methods can keep more details (clearer feather outline).

\section{Conclusion}
In this paper, we incorporate transformers and CNN to propose an efficient parallel transformer-CNN mixture block that utilizes the local modeling ability of CNN and the non-local modeling ability of transformers. Then, a new image compression architecture is designed based on the TCM block. Besides, we present a swin-transformer-based attention module to improve channel-wise entropy models. The results of experiments show that the image compression model with TCM blocks outperforms the CNN-only/Transformer-only based models under a suitable complexity. Furthermore, the performance of SWAtten surpasses previous attention modules designed for image compression. At last, our method achieves state-of-the-art on three different resolution datasets (i.e., Kodak, Tecnick, CLIC Professional Validation) and is superior to existing image compression methods.  

\section{Acknowledgment}
This paper is supported by Japan Science and Technology Agency (JST), under Grant JPMJPR19M5; Japan Society for the Promotion of Science (JSPS), under Grant 21K17770; Kenjiro Takayanagi Foundation; the Foundation of Ando Laboratory; NICT, Grant Number 03801, Japan.

{\small
\bibliographystyle{ieee_fullname}
\bibliography{main}
}

\clearpage
\appendix

\section{Classical Image Compression Standard Setting}
\subsection{VVC}
We use VTM-12.1 which is built form the website\footnote{\url{https://vcgit.hhi.fraunhofer.de/jvet/VVCSoftware_VTM/-/releases/VTM-12.1}} to achieve VVC. The script from CompressAI \footnote{\url{https://github.com/InterDigitalInc/CompressAI/tree/master/compressai/utils/bench}} is utilized to evaluate model. The command is as the following:
\begin{lstlisting}
    python -m compressai.utils.bench vtm [path of image folder];
    -c [path of VTM folder]/cfg/encoder_intra_vtm.cfg
    -b [path of VTM folder]/bin
    -q 16, 18, 20, 22, 24, 26, 28, 30, 32, 34, 36, 38, 40
\end{lstlisting}
\subsection{WebP}
We use the API of Pillow (PIL) to achieve WebP algorithm. The code is:
\begin{lstlisting}
    img.save(REC_WEBP, 'webp', quality=quality)
\end{lstlisting}
where quality is set as \{5,10,15,20,25,30,35,40,45,50\}.

\begin{table}[h]
\caption{BD-rate improvements against the VVC anchor. For different datasets, the anchor is recalculated based the corresponding dataset. Lower BD-rate represents higher performance.}
\setlength{\tabcolsep}{5mm}\begin{tabular}{|l|c|l|}
\hline
\textbf{Methods} & \multicolumn{1}{l|}{\textbf{Dataset}}                                        & \textbf{BD-Rate} \\ \hline \hline
Cheng \etal \cite{cheng2020learned}    & \multirow{7}{*}{\begin{tabular}[c]{@{}c@{}}Kodak\\ 768x512\end{tabular}}     & 3.16             \\ 
Xie \etal \cite{xie2021enhanced}        &                                                                              & -1.65             \\ 
Chen \etal \cite{chen2022two}     &                                                                              & -6.21            \\
He \etal \cite{he2022elic}       &                                                                              & -7.49            \\
\textbf{Ours (Large)}    &                                                                              & \textbf{-12.30}  \\ 
\textbf{Ours (Medium)}    &                                                                              & \textbf{-9.65}  \\ 
\textbf{Ours (Small)}    &                                                                              & \textbf{-7.39}  \\ 
\hline \hline
Ball\'{e} \etal \cite{balle2018variational}        & \multirow{5}{*}{\begin{tabular}[c]{@{}c@{}}Tecnick\\ 1200x1200\end{tabular}} & 30.66           \\
Xie \etal \cite{xie2021enhanced}     &                                                                              & -4.07             \\
Kim \etal \cite{kim2022joint}     &                                                                              & 6.98             \\
\textbf{Ours (Large)}    &                                                                              & \textbf{-13.71}   \\
\textbf{Ours (Medium)}    &                                                                              & \textbf{-11.29}   \\
\textbf{Ours (Small)}    &                                                                              & \textbf{-9.53}   \\
\hline \hline
Cheng \etal \cite{cheng2020learned}     & \multirow{7}{*}{\begin{tabular}[c]{@{}c@{}}CLIC-P val\\ 2K\end{tabular}}     & 6.77             \\ 
Xie \etal \cite{xie2021enhanced}        &                                                                              & -2.60            \\
Chen \etal \cite{chen2022two}      &                                                                              & -7.15            \\ 
Zou \etal \cite{zou2022devil}     &                                                                              & -3.68            \\ 
\textbf{Ours (Large)}    &                                                                              & \textbf{-11.85}   \\ 
\textbf{Ours (Medium)}    &                                                                              & \textbf{-10.27}   \\
\textbf{Ours (Small)}    &                                                                              & \textbf{-8.94}   \\
\hline \hline
\textbf{VVC}     & -                                                        & 0                \\ \hline
\end{tabular}
\label{bdrate}
\end{table}
\section{Detailed Network Architecture}

\begin{figure*}
\centering
\includegraphics[scale=0.8]{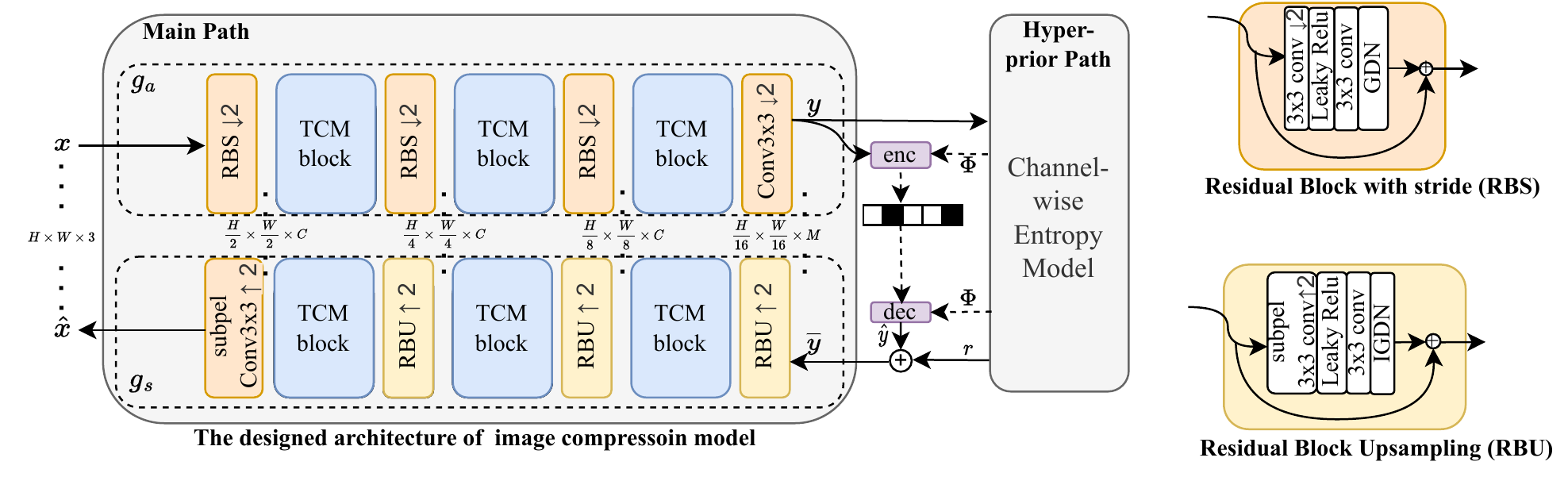}
		\caption{The overall framework (left). The architectures of RBS and RBU in \cite{cheng2020learned} (right).}
		\label{fig:33}
\end{figure*}

The architecture of the our method is shown in Fig. \ref{fig:33}. The head dimensions of TCM blocks in $g_a$ and $g_s$ are set as \{8, 16, 32, 32, 16, 8\}, while the head dimensions of TCM blocks in $h_a$ and $h_s$ are set as 32. We set channel numbers $C$ of TCM blocks as 128/192/256 for Small/Medium/Large model. RBS and RBU have the same architectures as in \cite{cheng2020learned}. The numbers of channels of the middle convolutional layers in RBS and RBU are 64/96/128 for our Small/Medium/Large model, while the number of last layer of RBS and RBU is 128/192/256. Here, to achieve the balance between running speed and RD-performance, we reduce the slices number in \cite{minnen2020channel} from 10 to 5. Therefore, we have 5 Channel-Conditional Parameter Nets with SWAtten to get \{$\boldsymbol{\mu_0}, \boldsymbol{\mu_1}, \boldsymbol{\mu_2}, \boldsymbol{\mu_3}, \boldsymbol{\mu_4}$\} and \{$\boldsymbol{\sigma_0}, \boldsymbol{\sigma_1}, \boldsymbol{\sigma_2}, \boldsymbol{\sigma_3}, \boldsymbol{\sigma_4}$\}. Also, we have 5 Latent Residual Prediction to get \{$\boldsymbol{\overline{y}_0}, \boldsymbol{\overline{y}_1}, \boldsymbol{\overline{y}_2}, \boldsymbol{\overline{y}_3}, \boldsymbol{\overline{y}_4}$\}. All the restored slices are concatenated as $\boldsymbol{\overline{y}}$ which is sent to decoder $g_s$ to get a decompressed image. 

\section{Comparison with Recent LIC Works}
To get quantitative results, we present the BD-rate \cite{bdrate} computed from PSNR-BPP curves as the quantitative metric. The anchor RD-performance is set as the results of VVC on different datasets (BD-rate=0\%). The Table \ref{bdrate} shows the results. As results show, we outperform the previous works and achieve SOTA performance based on the three datasets with different resolutions.

\section{Ablation Studies on Various Entropy Estimation Models}
\begin{figure}[h!]
    \centering\includegraphics[scale=0.5]{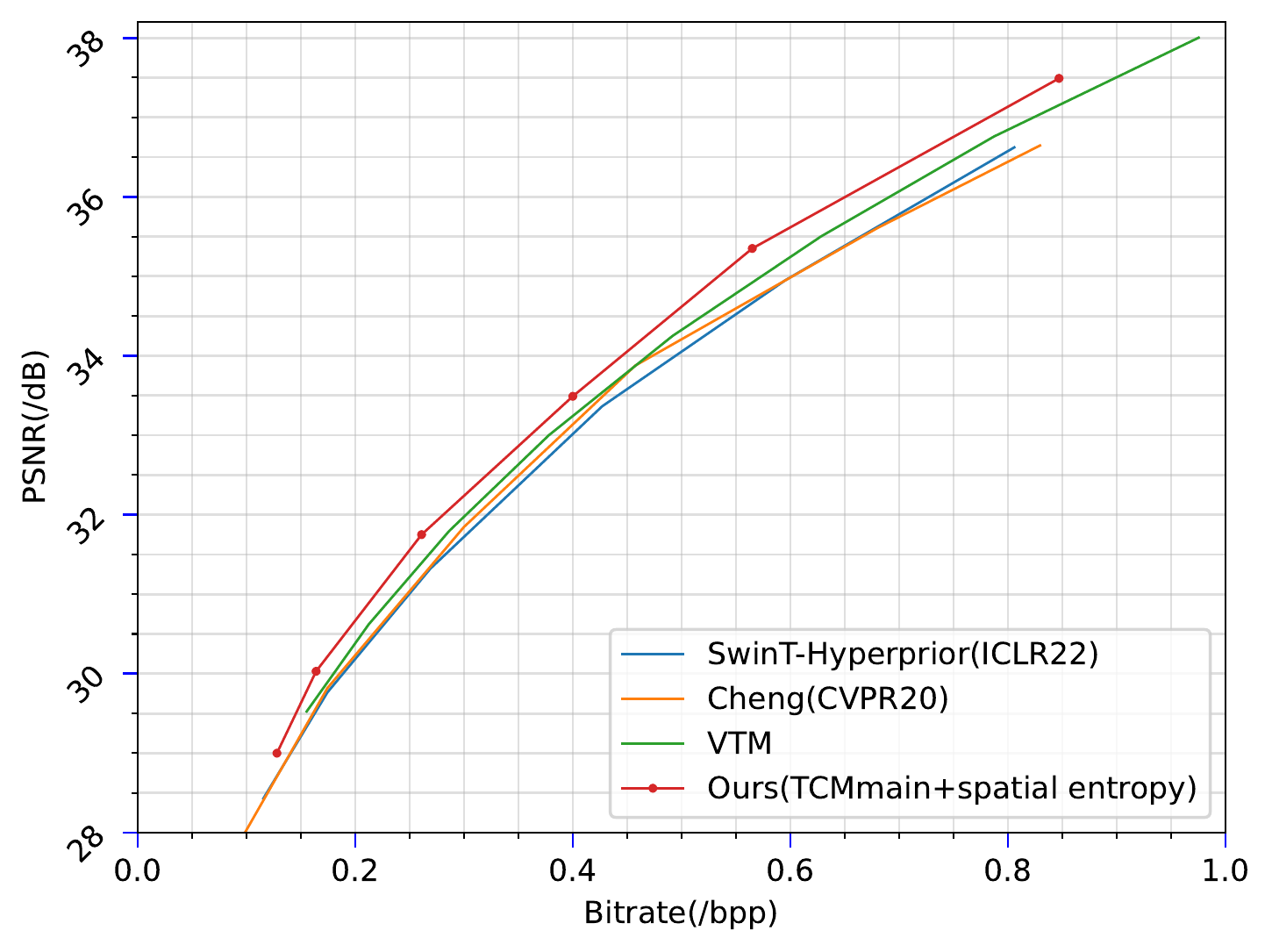}
  \caption{Performance evaluation of the models using the spatial-wise entropy model \cite{minnen2018joint} on the Kodak dataset.}
  \label{fig:3}
\end{figure}

To verify our TCM blocks can improve the overall RD-performance, in addition to test the model with the channel-wise entropy model in \cite{minnen2020channel}, we also try the model using the spatial-wise entropy model in \cite{minnen2018joint}. The results are shown in Fig. \ref{fig:3}. We define the model where the main path uses TCM block as ``TCMmain". We compare the TCMmain model using spatial-wise entropy model with ``SwinT-Hyperprior" model in \cite{zhu2022transformerbased} and the model in \cite{cheng2020learned}. All of these three methods use spatial-wise entropy models. The difference is that our model is based on TCM block, the model in \cite{cheng2020learned} is based on CNN, and ``SwinT-Hyperprior" is based on swin-transformer. As we can see, our method can get the best RD-performance. This suggests that the TCM blocks can significantly improve image compression, and are robust to different entropy models.
\section{Ablation Studies on the Numbers of Slices}
\begin{figure}[h!]
    \centering\includegraphics[width=.9\linewidth]{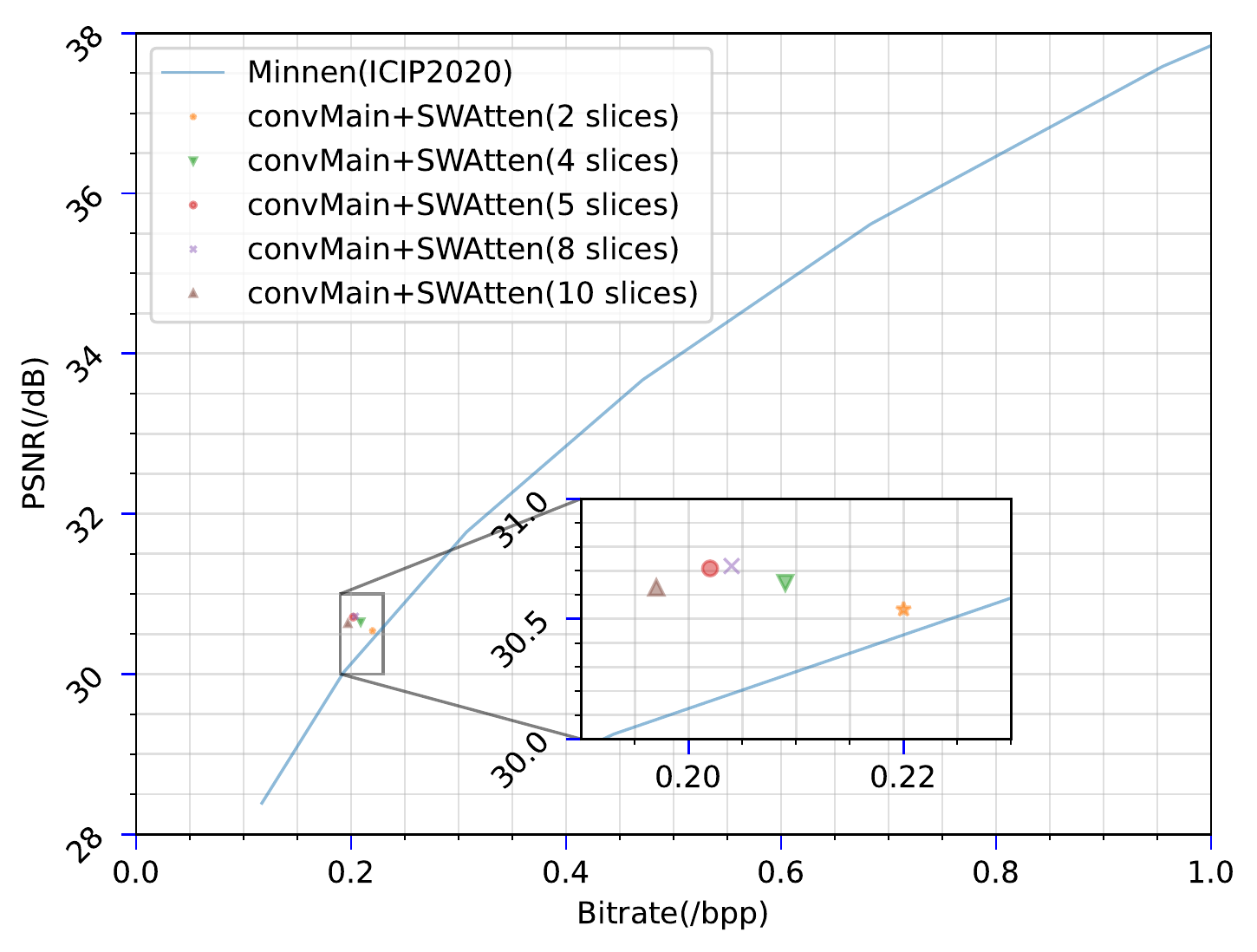}
  \caption{Effect of different slices numbers on RD-performance. ``convMain" means we use the main path in \cite{minnen2020channel}.}
  \label{fig:10}
\end{figure}

The number of slices $s$ is an important hyper-parameter for channel-wise entropy model in \cite{minnen2020channel}. A larger number leads to lower efficiency, while a lower number causes a worse RD-performance. To find a suitable number, we test some different number setting $s=\{2,4,5,8,10\}$ for our entropy model with the proposed SWAtten. The main path of the tested model is the same as the main path in \cite{minnen2020channel}. The entropy model is also similar, the difference is that we add SWAtten. The results are shown in Fig. \ref{fig:10}. As we can see, when $s$ is low, we get a bad RD-performance. With $s$ increasing, the performance is improved. But when $s>5$, 
the improvement of the performance is not obvious, and even decreases. This indicates that 5 slices have been able to learn enough information in our model. Therefore, we set $s$ as 5 in our model to achieve the balance between running speed and RD-performance.

\section{Abaltion Studies on the Design of SWAtten}
\par
\begin{figure}[h!]
        \centering
        \includegraphics[scale=0.3]{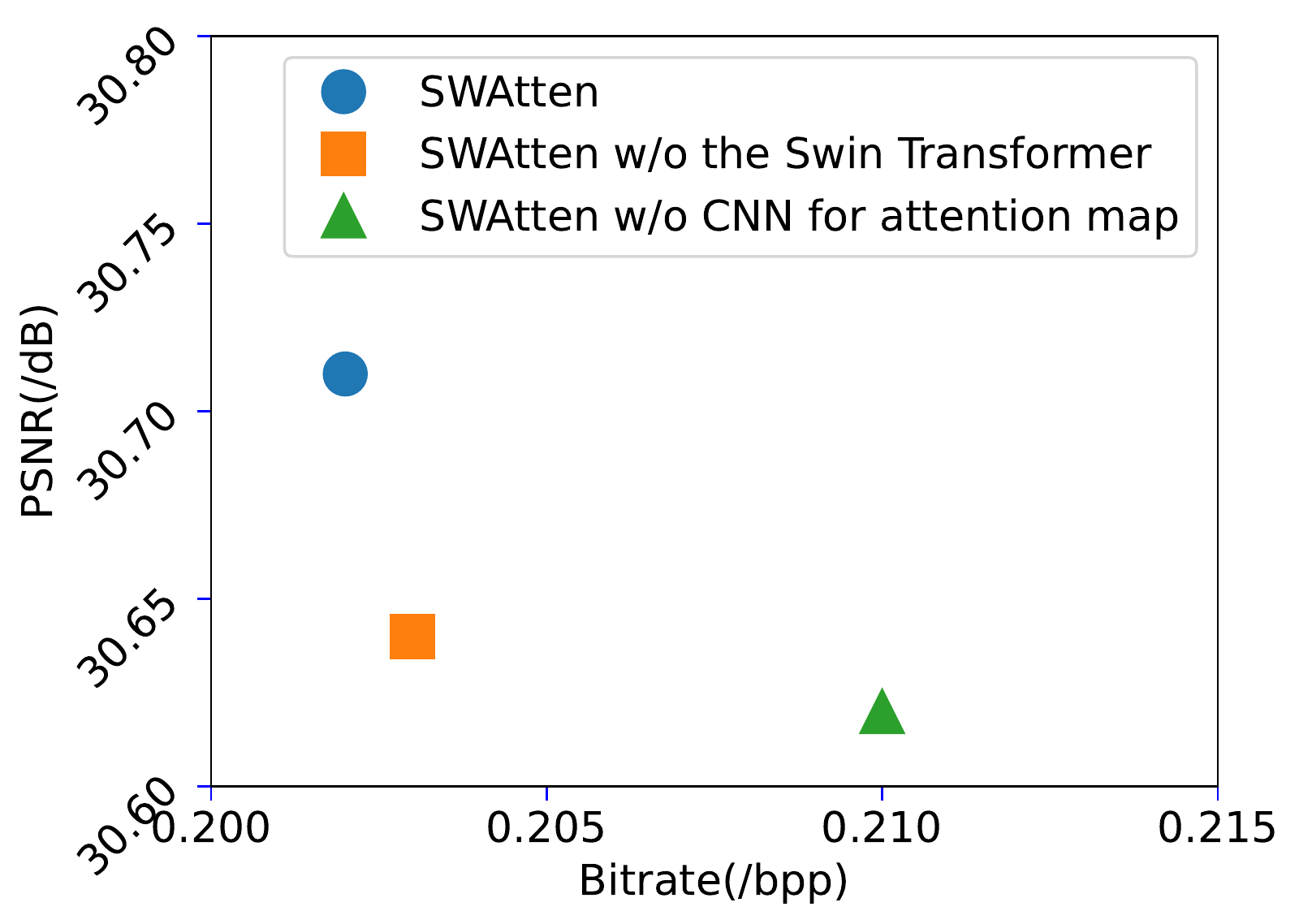}
        \caption{The ablation study on SWAtten (using the same main path as \cite{minnen2020channel}.}
        \label{fig:b2}
\end{figure}

We test the SWAtten w/o CNN for attention map (green point) in Fig. \ref{fig:b2}. In addition, we also evaluate the case w/o the swin transformer (yellow point). From the results, we can see that using either of these two modules can bring about 0.1dB PSNR improvement with fewer bitrates.

\section{Visualization}
\begin{figure*}[h!]
    \centering\includegraphics[scale=0.5]{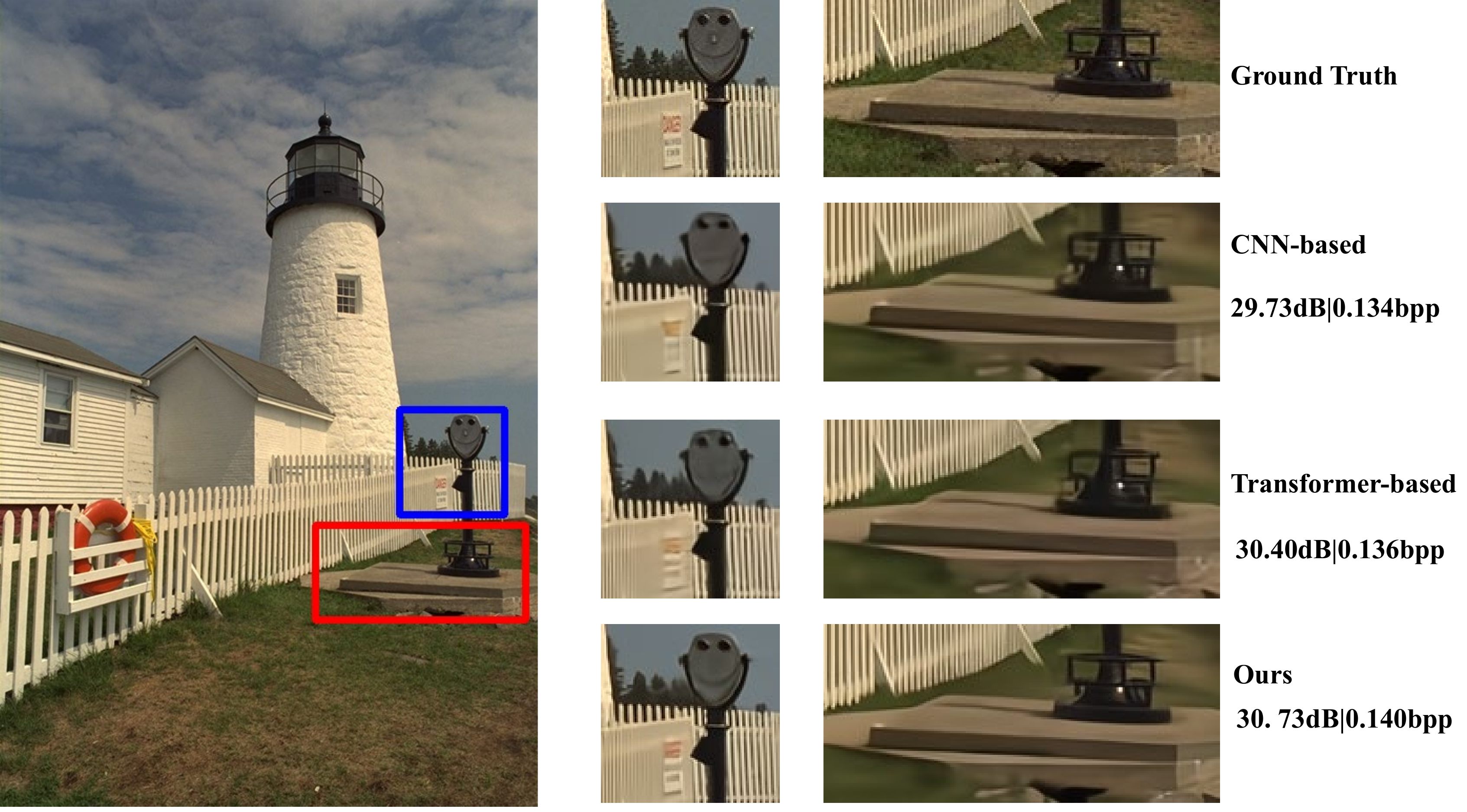}
  \caption{The visualization of $kodim19$ in Kodak dataset by using our TCM-based model, CNN-based model \cite{cheng2020learned}, and transformer-based model \cite{zou2022devil}. PSNR$|$Bit-rate is listed in the last column.}
  \label{fig:4}
\end{figure*}

We conducted a comparison between our TCM-based model and both a CNN-based \cite{cheng2020learned} and Transformer-based \cite{zou2022devil} model using the Kodak dataset's $kodim19$ and $kodim20$ images. The results of this comparison are presented in Fig. \ref{fig:4}. We focused our analysis on two local regions, and the differences between the three models are noticeable. In the upper local area of $kodim19$, our method effectively reconstructed some of the road sign details while making it easier to differentiate between the back fences. The pasted poster color is also clearer and not distorted. In contrast, our method generated fewer artifacts compared to the CNN-based/Transformer-based models when reconstructing the lower local region. For $kodim20$, our method generated the clearest sign in the left local region, and we achieved a clearer ``E" letter than the other two methods in the right local region.

\begin{figure*}[h!]
    \centering\includegraphics[scale=0.5]{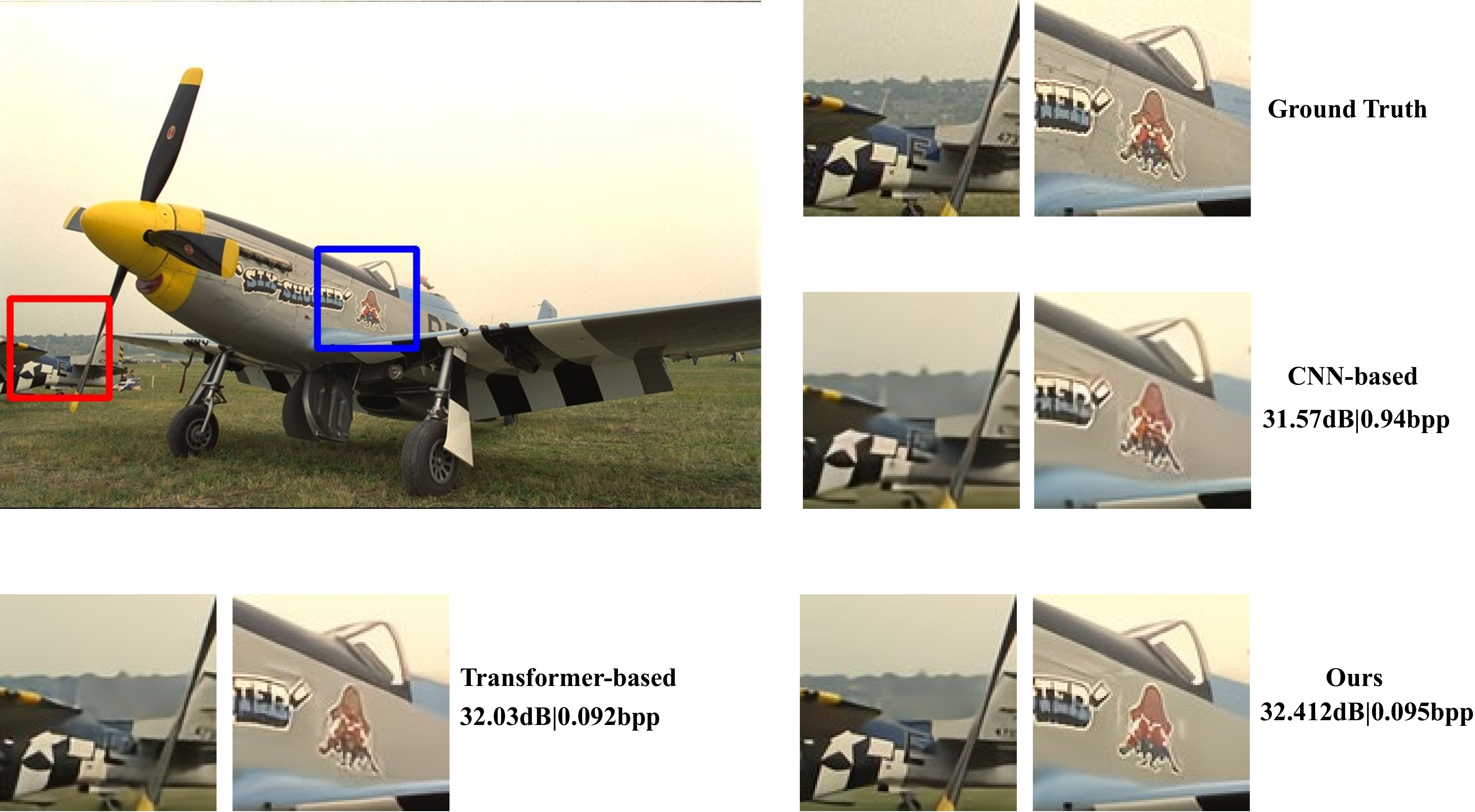}
  \caption{The visualization of $kodim20$ in Kodak dataset by using our TCM-based model, CNN-based model \cite{cheng2020learned}, and transformer-based model \cite{zou2022devil}. PSNR$|$Bit-rate is listed on the subfigures' right.}
  \label{fig:11}
\end{figure*}

\end{document}